\documentclass[sigplan,10pt,review,acmsmall]{acmart}
\AtBeginDocument{%
  \providecommand\BibTeX{{%
    \normalfont B\kern-0.5em{\scshape i\kern-0.25em b}\kern-0.8em\TeX}}}

\setcopyright{none}
\copyrightyear{2021}
\acmYear{2021}

\acmJournal{JACM}

\newcommand{\ifdebug}{false}
\newcommand{\rsource}{arxiv}

\usepackage[normalem]{ulem}
\ifnum\pdfstrcmp{\ifdebug}{true}=0
\else
\renewcommand\sout[1]{}
\fi
\newcommand{\debugte}[2]{%
  \ifnum\pdfstrcmp{\ifdebug}{true}=0
  #1%
  \else
  #2
  \fi
}

\newcommand{\releasesource}[0]{%
  \ifnum\pdfstrcmp{\rsource}{arxiv}=0
  in the appendix
  \else
  in the supplementary material
  \fi
}

\usepackage{booktabs}   
\usepackage{subcaption} 
\usepackage{subfiles}
\usepackage{amsmath}
\usepackage{microtype}
\usepackage{multirow}
\definecolor{DOrange}{RGB}{145,20,20}
\definecolor{Red}{RGB}{255,0,0}
\definecolor{DeepGreen}{RGB}{5,102,8}

\newcommand\NV[1]{\debugte{\textcolor{magenta}{#1}}{#1}} 
\renewcommand\C[1]{\mathtt{#1}}
\renewcommand\S[1]{\mathsf{#1}}
\newcommand\I[1]{\mathit{#1}}
\newcommand\unitaryc[1]{\llbracket \mathit{#1} \rrbracket}

\newcommand{\name}{{\sf Elrond}}
\usepackage{setspace}
\usepackage[ruled,linesnumbered,vlined]{algorithm2e}
\SetAlFnt{\small\sffamily}
\SetKwRepeat{Do}{do}{while}
\SetKwRepeat{Repeat}{repeat}{until}
\SetKw{Break}{break}
\SetKw{ZFor}{for}
\SetKw{ZDo}{do}
\SetKw{ZForall}{forall}
\SetKw{UntilConverge}{\NV{until converge}}
\SetKw{Continue}{continue}
\SetKwProg{Match}{match}{\string:}{}
\SetKwInOut{All}{all}
\SetKwInOut{Params}{Inputs}
\SetKwInOut{Output}{Output}
\SetKwProg{Procedure}{Procedure}{}{}
\usepackage{graphicx}
\usepackage{hyperref}
\hypersetup{
    colorlinks=true,
    linkcolor=blue,
    filecolor=magenta,
    urlcolor=cyan,
}

\newcommand*{\fullref}[1]{\hyperref[{#1}]{\autoref*{#1}}}
\usepackage{listings}
\definecolor{codegreen}{rgb}{0,0.6,0}
\definecolor{codegray}{rgb}{0.5,0.5,0.5}
\definecolor{codepurple}{rgb}{0.58,0,0.82}
\definecolor{backcolour}{rgb}{1.00,1.00,1.00}
\definecolor{CBOrange}{RGB}{210, 77, 0}
\definecolor{CBLightOrange}{RGB}{255, 147, 60}
\definecolor{CBPurple}{RGB}{93, 58, 155}

\lstdefinelanguage{ocaml}[Objective]{Caml}{
  deletekeywords={closed,ref},
  morekeywords={initializer},
}

\lstdefinestyle{mystyle}{
    frame = single,
    backgroundcolor=\color{backcolour},
    commentstyle=\color{codegreen},
    keywordstyle=\color{magenta},
    numberstyle=\tiny\color{codegray},
    stringstyle=\color{codepurple},
    basicstyle=\ttfamily\footnotesize,
    breakatwhitespace=false,
    breaklines=true,
    captionpos=b,
    keepspaces=true,
    numbers=left,
    numbersep=5pt,
    showspaces=false,
    showstringspaces=false,
    showtabs=false,
    tabsize=2,
    framexleftmargin=8pt
}

\usepackage{tikz}
\newcounter{markeq}
\newcommand{\pstrut}[1]{\vrule height0pt depth0pt width0pt #1 \fboxsep}
\newcommand*\bmarkeq{\stepcounter{markeq}%
  \tikz[remember picture]\node(startframe-\themarkeq){\pstrut{height}};%
  \kern\fboxsep}
\newcommand*\emarkeq[1]{\kern\fboxsep
  \begin{tikzpicture}[remember picture,overlay]
    \node (endframe-\themarkeq){\pstrut{depth}};
    \draw[,#1,opacity=0.8] (startframe-\themarkeq.north)
      rectangle (endframe-\themarkeq.south);
  \end{tikzpicture}%
}

\begin{document}


\title{Data-Driven Abductive Inference of Library Specifications}

\author{Zhe Zhou}
\affiliation{
  \institution{Purdue University}            
  \country{USA}                    
}

\author{Robert Dickerson}
\affiliation{
  \institution{Purdue University}            
  \country{USA}                    
}

\author{Benjamin Delaware}
\affiliation{
  \institution{Purdue University}            
  \country{USA}                    
}

\author{Suresh Jagannathan}
\affiliation{
  \institution{Purdue University}            
  \country{USA}                    
}

\renewcommand{\shortauthors}{Zhou et. al}

\begin{abstract}
  Programmers often leverage data structure libraries that provide
  useful and reusable abstractions.  Modular verification of programs
  that make use of these libraries naturally rely on specifications
  that capture important properties about how the library expects
  these data structures to be accessed and manipulated.  However,
  these specifications are often missing or incomplete, making it hard
  for clients to be confident they are using the library safely.  When
  library source code is also unavailable, as is often the case, the
  challenge to infer meaningful specifications is further exacerbated.
  In this paper, we present a novel data-driven abductive inference
  mechanism that infers specifications for library methods sufficient
  to enable verification of the library's clients.  Our
  technique combines a data-driven learning-based framework to
  postulate candidate specifications, along with SMT-provided
  counterexamples to refine these candidates, taking special care to
  prevent generating specifications that overfit to sampled tests.
  The resulting specifications form a minimal set of requirements on
  the behavior of library implementations that ensures safety of a
  particular client program. Our solution thus provides a new
  multi-abduction procedure for precise specification inference of
  data structure libraries guided by client-side verification tasks.
  Experimental results on a wide range of realistic OCaml data
  structure programs demonstrate the effectiveness of the approach.
\end{abstract}



\maketitle

\section{Introduction}
\label{sec:intro}
Using a specification of a library's methods in the verification of
its clients is a hallmark of modular reasoning.  Because these
specifications encapsulate the interface between the client and the
library, each may be independently verified without
access to the other's implementation.  This modularity
is particularly beneficial when the library function is complex or its
source code is unavailable. All too often, though, such specifications
are either missing or incomplete, preventing the verification of clients
without making (often unwarranted) assumptions about the
behavior of the library.  This problem is further
exacerbated when libraries expose rich datatype functionality, which
often leads to specifications that rely on inductive
invariants~\cite{DilligAbductiveInv,itzhaky2014modular}
and complex structural relations.  One solution to this problem is to
automatically \emph{infer} missing specifications. Unfortunately,
while significant progress has been made in specification inference
over the past several
years~\cite{AlbarghouthiMSS,zhu2016automatically,padhi2016data,RepInvGen},
existing techniques have not considered inference in the frequently
occurring case of client programs that make use of data structure
libraries with unavailable implementations.

To highlight the challenge, consider the following simple program,
which concatenates two stacks together using four operations provided
by a \textsf{Stack} library: $\S{push}$,
$\S{top}$, $\S{is\_empty}$ and $\S{tail}$.
\begin{lstlisting}[language = ML,linewidth=350pt, xleftmargin=50pt]
let rec concat s1 s2 =
  if Stack.is_empty s1 then s2
  else Stack.push (Stack.top s1) (concat (Stack.tail s1) s2)
\end{lstlisting}
\noindent
To ensure the correctness of this client function, its author may wish
to verify that (a) the top element of the output stack is always the
top element of one of the input stacks; and, (b) every element of the
output stack is also an element of one of the input stacks and
vice-versa. In order to express this behavior in a form amenable for
automatic verification, we need some mechanism to encode the semantics
of stacks in a decidable logic. To do so, we rely on a pair of
\emph{method predicates}, "$a$ is the head of stack $s$", $hd(s, a)$,
and "$a$ is a member of stack $s$", $mem(s, a)$, to write our postcondition:
\begin{align}
  \forall u, &(\I{hd}(\nu, u) \implies \I{hd}(\C{s}_1, u) \lor
               \I{hd}(\C{s}_2, u))\ \land (\I{mem}(\nu, u) \iff
               \I{mem}(\C{s}_1, u) \vee \I{mem}(\C{s}_2, u)) \tag{$\phi_{\C{concat}}$}
\end{align}
We assume these method predicates are associated with (possibly
blackbox) implementations that we can use to check the specifications
in which they appear. For example, $\I{hd}$ may be defined in terms of
the stack operations $\S{top}$ and $\S{is\_empty}$, while the
implementation of $\I{mem}$ might additionally use $\S{tail}$.  The variable $\nu$
in $\phi_{\C{concat}}$ is used to represent the output stack of $\C{concat}$.  The above assertion
claims that the head of the output stack must be the head of either
$\C{s}_1$ or $\C{s}_2$ and that any element found in the output must be
a member of either $\C{s}_1$ or $\C{s}_2$.


By treating method predicates ($\I{hd}$ and $\I{mem}$) and library functions
($\S{push}$, $\S{top}$, $\S{is\_empty}$, and $\S{tail}$) as uninterpreted
function symbols, it is
straightforward to generate verification conditions (VCs), e.g.
using weakest precondition inference, which can be handed off to an
off-the-shelf SMT solver like Z3 to check.  However, the counter-examples
returned by the theorem prover may be spurious, generated by incorrect
assumptions about library method behavior in the absence of any
constraints on these behaviors outside the client VCs.  For example,
the prover might assume the formula $\neg\I{hd}(\C{s}_1,
\S{top}(\C{s}_1))$ is valid, i.e. that the
result of $\S{top}(\C{s}_1)$ is not the head of $\C{s}_1$. This claim
is obviously inconsistent with the client's expectation of $\S{top}$'s semantics, but it is not disallowed
by any constraints in the SMT query.  Using
this assumption, Z3 may return the following counterexample:
$\exists \I{s\;u}. \I{hd}(\I{s,u}) \land \lnot \I{mem}(\I{s},\I{u})$.
This counterexample, which is obviously incongruous with the
intended semantics of $\I{hd}$ and $\I{mem}$, occurs because
the expected relationship between \emph{hd} and \emph{mem} is lost
when the predicates are embedded as uninterpreted functions in the
SMT query.

To overcome this problem, we need stronger specifications for the
library methods, defined in terms of these method predicates, that are
sufficient to imply the desired client postcondition. In particular,
these specifications should rule out spurious unsafe executions such
as the counter-example given above.  In the (quite likely) scenario
that such library specifications are not already available, a
reasonable fallback is to infer \emph{some} specifications for these
functions that are strong enough to ensure the safety of the client.
Traditional approaches to specification inference usually adopt a
closed-world assumption in which specifications of library methods are
discovered in isolation, independent of the client context in which
they are being used.  Such assumptions are not applicable here since
(a) we do not have access to the library's method implementations and
(b) the nature of the specifications we need to infer are impacted by
the verification demands of the client.  In this setting, some form of
data-driven inference~\cite{RepInvGen, padhi2016data,
  zhu2016automatically} can be beneficial. Such an approach may be
tailored to the client context in which the library methods are used,
postulating candidate specifications for library methods based on
observations of their input-output behavior. Unfortunately, completely
blackbox data-driven approaches are susceptible to overfitting on the
set of observations used to train them, and can thus discount
reasonable and safe behaviors of the underlying library functions.



To address the problem of overfitting, we might instead consider
attacking this problem from a purely logical standpoint, treating
specification inference as an instance of a multi-abductive
inference problem~\cite{AlbarghouthiMSS} that tries to find formulae
$\S{R_{push}}$, $\S{R_{top}}$, $\S{R_{tail}}$ and $\S{R_{is\_empty}}$
such that $\bigwedge R_i \not \models \bot$ and yet which are sufficient to
prove the desired verification condition.  While such problems have
been previously solved over linear integer arithmetic
constraints~\cite{AlbarghouthiMSS} using quantifier elimination, these
prior techniques cannot be directly applied to formulae with
uninterpreted function symbols like the method predicates
(e.g., $\I{hd}$ and $\I{mem}$) used to encode library method
specifications in our setting.

In this work, we combine aspects of these data-driven and abductive
approaches in a way that addresses the limitations each approach has
when considered independently.  Our technique uses SMT-provided
counterexamples to generate infeasible interpretations of these
predicates (similar to other abductive inference methods) while using
concrete test data to generate feasible interpretations (similar to
data-driven inference techniques).  This combination yields a novel
CEGIS-style inference methodology that allows us to postulate
specifications built from method predicates sufficient to prove the
postcondition in a purely blackbox setting. \NV{The specifications
  learned by this procedure are guaranteed to be both
  \emph{consistent} with the observed input-output behavior of the
  blackbox library implementations and \emph{safe} with respect to
  the postcondition of the client program.} As there may be many such
specifications, we also endeavor to find a \NV{\emph{maximal} one}
that is at least as weak as every other safe and consistent
specification, in order to avoid overfitting to observed library
behaviors.  Our algorithm applies another data-driven weakening
procedure to find these maximal specifications.

\begin{figure}[!t]
  \includegraphics[width=400pt]{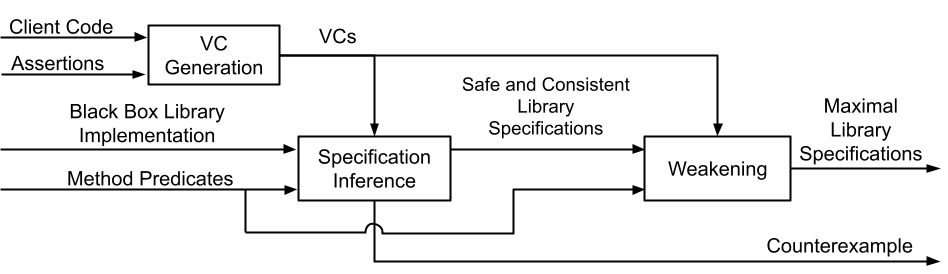}
\centering
\caption{\small \name\ pipeline.
    }
\label{fig:workflow_simplified}
\end{figure}

To demonstrate the effectiveness of our approach, we have implemented
a fully automated abductive specification inference pipeline in OCaml
called \name\ (see \autoref{fig:workflow_simplified}).  This pipeline
takes as input (a) an OCaml client program that may call blackbox
library code defined over algebraic datatypes like lists, trees,
heaps, etc.; (b) assertions about the behavior of this client program;
and, (c) a set of method predicates (e.g., \emph{hd} or \emph{mem}),
along with their (possibly blackbox) implementations, that are used to
synthesize library method specifications.
It combines tests and counterexample-guided refinement techniques to
either generate a set of \NV{maximal}
specifications for the library methods used by the client program, or a
counterexample that demonstrates a violation of
the postcondition.  The notion of ``weakest'' used in our definition
of maximal is bounded by the ``shape'' of specifications (e.g., the
number of quantified variables, the set of method predicates, etc.)
and a time bound.  Our results over a range of sophisticated
data-structure manipulating programs, including those drawn from e.g.,
~\citet{okasaki1999purely}, show that \name{} is able to discover
maximally-weak specifications (as determined by an oracle executing
without any time constraints) for the vast majority of applications in
our benchmark suite within one hour.

Our key contribution is thus a new abductive inference framework that
is a fusion of automated data-driven methods and counterexample-guided
refinement techniques, tailored to specification inference for
libraries that make use of rich algebraic datatypes.  Specifically,
we:

\begin{description}
\item[1.]  Frame client-side verification as a multi-abduction
  inference problem that searches for library method specifications
  that are both consistent with the method's implementation and
  sufficient to verify client assertions.
\item[2.] Devise a novel specification weakening procedure that yields
  the weakest specification among the collection of all safe and
  consistent ones with respect to a given set of quantified variables
  and method predicates.
\item[3.]  Evaluate our approach in a tool, \name{}, which we use to
  analyze a comprehensive set of realistic and challenging functional
  (OCaml) data structure programs. An artifact containing this tool and our benchmark suite is publicly available~\cite{zhe_zhou_2021_5130646}.
\end{description}

\noindent The remainder of the paper is structured as follows.  The
next section presents \NV{an overview of our approach using a detailed
example to motivate its key ideas}. A formal characterization of the
problem is given in \fullref{sec:formalization}.
~\fullref{sec:learning} defines how a data-driven learning
strategy can be used to perform inference.  A detailed presentation of
the algorithm used to manifest these ideas in a practical
implementation is given in \fullref{sec:algorithm}.
Details of our implementation and evaluation results are explained in
\fullref{sec:evaluation}.  Related work and conclusions are given in
\fullref{sec:related} and \fullref{sec:conc}.

\section{Overview and Motivation}
\label{sec:motivation}

\NV{We divide the inference of maximal library specifications into two
  stages, which are represented as the ``Specification Inference'' and
  ``Weakening'' components in \autoref{fig:workflow_simplified}. Both
  stages leverage data-driven learning to overcome the lack of a
  purely logical abduction procedure for our specification language.
  The initial inference stage learns a set of safe and consistent
  specifications from a combination of concrete tests and
  verifier-provided counterexamples. The next stage then weakens these
  specifications by iteratively augmenting this data set with
  additional safe behaviors until a set of maximal specifications are
  found.
}

\begin{figure}[ht]
\includegraphics[width=400pt]{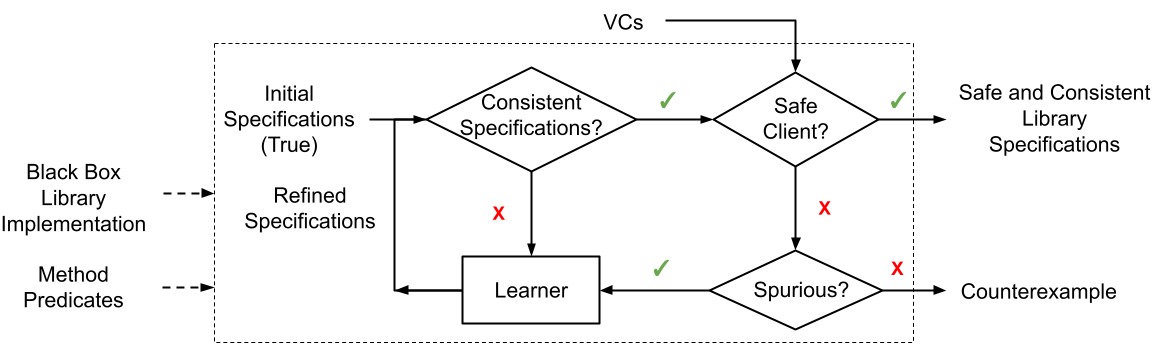}
\centering
\caption{\small The details of \name{}'s Specification Inference component
  from \autoref{fig:workflow_simplified}. The blackbox library
  implementation and method predicates are used by multiple
  components; their corresponding arrows within the diagram are omitted for
  clarity.
    }
\label{fig:specinfer}
\end{figure}

\NV{\autoref{fig:specinfer} provides a more detailed depiction of the
  initial specification inference stage in \name{}. Starting from an
  initial set of maximally permissive specifications, this stage
  iteratively refines the set of candidate specifications until either
  a set of safe and consistent solutions or a counterexample
  witnessing an unsafe execution is found.}

  Each iteration first uses a property-based sampler, e.g.
  QuickCheck~\cite{claessen2011quickcheck}, to look for executions of
  the blackbox library implementations that are inconsistent with the
  current set of inferred specifications.  \NV{The reliance on a
  generator to provide high-quality tests provides yet more motivation
  for the subsequent weakening phase, in order to ensure that the final
  abduced specifications are not overfitted to or otherwise biased by
  the tests provided by the generator.  At the same time, we also observe that the sorts of
  shape properties (e.g., membership and ordering) used in our
  specifications and assertions are relatively under-constrained and
  are thus amenable to property-based random sampling.  We do not ask,
  for example, for QuickCheck to generate inputs satisfying
  non-structural properties like ``\emph{a list whose 116$^{th}$
    element is equal to 5.''}}

 \NV{Any tests that are disallowed by the current solution are passed to
  a learner which uses them to generalize the current specification.
  If no inconsistencies are detected, \name{} attempts to verify the
  client against the candidate specifications using a theorem
  prover. If the inferred specifications are sufficient to prove
  client safety, the loop exits, returning the discovered solution. If
  not, the verifier has identified a model that represents a
  \emph{potential} safety violation. The model is then analyzed in an
  attempt to extract test inputs that trigger a safety violation. If
  we are unable to find such a counterexample, the model is most
  likely incongruous with the semantics of the method predicates and
  is thus spurious. In this case, the model is passed to the learner
  so that it can be used to strengthen candidate specifications, preventing this
  and similar spurious counterexamples from manifesting in subsequent iterations.}

\begin{figure}[ht]
\includegraphics[width=350pt]{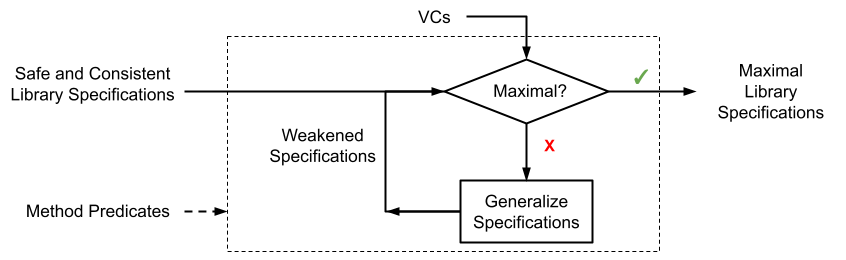}
\centering
\caption{\small The details of \name{}'s Weakening component from
  \autoref{fig:workflow_simplified}. Arrows for method predicates are
  again omitted for clarity.}
\label{fig:weakening}
\end{figure}

\NV{While the previous loop is guaranteed to return safe and
  consistent solutions, it may find specifications that are nonetheless \emph{too
    strong} with respect to the underlying library
  implementation. This occurs when the property-based sampler fails to
  find a test that identifies an inconsistent specification, which may
  happen when the input space of a library function is very large. To
  combat overfitting specifications to test data, candidate solutions
  are iteratively weakened using the data-driven counterexample-guided
  refinement loop depicted in \autoref{fig:weakening}.  The
  data in this phase is supplied by the underlying theorem prover
  rather than a concrete test generator.  Each iteration of the
  refinement loop first attempts to find a safe execution of the
  client program that is disallowed by the current set of
  specifications. If no such execution can be found, the
  specifications are maximal and the loop terminates. Otherwise, the
  identified execution is passed to a learner, which uses it to
  generalize the candidate solution so that the execution is permitted
  before continuing the refinement loop. The learner always generalizes
  candidate specifications, maintaining the invariant that the current
  solution is consistent with all previously observed library behaviors.}

\subsection{\name{} in action}
\label{subsec:motivation-problemsetup}
\NV{To illustrate our approach in more detail, we apply it to the
  stack concatenation example from the introduction.
  Given the postcondition $\phi_{\C{concat}}$ and the implementation
  of $\C{concat}$ from \autoref{sec:intro}, \name{} generates a formula
  that can be simplified to the following implication:}

\begingroup \setlength{\jot}{2pt}
  \begin{align*}
  \forall &\C{s}_1, \C{s}_2, \nu, \nu_{\S{top}}, \nu_{\S{tail}}, \nu_{\S{concat}}, \nu_{\S{is\_empty}},\\
  (       &\textcolor{CBOrange}{\S{R_{is\_empty}}(\C{s}_1, \nu_{\S{is\_empty}})} \land
          \textcolor{CBOrange}{\S{R_{top}}(\S{s}_1, \nu_{\S{top}})}
            \land \textcolor{CBOrange}{\S{R_{tail}}(\C{s}_1,
            \nu_{\S{tail}})} \land
            \textcolor{CBOrange}{\S{R_{push}}(\nu_{\S{top}}, \nu_{\C{concat}}, \nu)})\\
          & \implies \\
    (       &\textcolor{CBPurple}{(\nu_{\S{is\_empty}} = \bot \land
              \forall u, (\I{hd}(\nu_{\S{concat}}, u)  \implies \I{hd}(\nu_{\S{tail}}, u) \lor \I{hd}(\C{s}_2, u)) \land} \\
          &\ \quad\quad \textcolor{CBPurple}{(\I{mem}(\nu_{\S{concat}}, u) \iff \I{mem}(\nu_{\S{tail}}, u) \vee \I{mem}(\C{s}_2, u))) \implies} \\
          &\ \textcolor{CBPurple}{\forall u, (\I{hd}(\nu, u) \implies \I{hd}(\C{s}_1, u) \lor \I{hd}(\C{s}_2, u))\ \land (\I{mem}(\nu, u) \iff \I{mem}(\C{s}_1, u) \vee \I{mem}(\C{s}_2, u))})
\end{align*}
\endgroup

\NV{The four predicates in the premise of this formula correspond to
  the four library functions ($\S{push}$, $\S{top}$, $\S{is\_empty}$
  and $\S{tail}$) invoked in a recursive call to \lstinline|concat|. The
  specification of a blackbox library function $\I{f}$ in our
  assertion logic is represented as a \emph{placeholder predicate}: an uninterpreted predicate that
  relates the parameters of $\I{f}$ to its return value. For a
  library function $\I{f}$, we adopt a naming convention of
  $\S{R_{f}}$ and $\nu_{\I{f}}$ for its placeholder predicates and return
  values, respectively. The predicate
  $\S{R_{top}}(\S{s}_1, \nu_{\S{top}})$ in the above formula, for
  example, says the variable $\nu_{\S{top}}$ holds the return
  value of the call to \lstinline|Stack.top s1| in
  \lstinline|concat|. The result of the recursive call to
  \lstinline|concat| is similarly denoted as $\nu_{\C{concat}}$. The
  conclusion of the formula encodes the expected verification
  condition for a recursive call to \lstinline|concat|: namely, that if the
  result of the call to \lstinline|Stack.is\_empty s1| is \textsf{false} and
  the recursive call to \lstinline|concat (Stack.tail s1) s2|
  satisfies $\phi_{\C{concat}}$, then the result of \lstinline|concat| must
  also satisfy $\phi_{\C{concat}}$. The remainder of this section
  refers to the premise and conclusion of this implication as
  $\textcolor{CBOrange}{\Sigma_{concat}}$ and
  $\textcolor{CBPurple}{\Phi_{\C{concat}}}$, respectively. }



\NV{\paragraph{Method Predicates} From a logical standpoint, the
  method predicates used in $\textcolor{CBPurple}{\Phi_{\C{concat}}}$
  are simply uninterpreted function symbols which have no intrinsic
  semantics.
  This representation allows our specifications to use
  predicates whose semantics may be difficult to encode directly in the
  logic.  Embedding recursively defined predicates like \lstinline|mem|, for
  example, requires particular care~\cite{zhu2016automatically}.
  In order to ensure
  that the specifications inferred by \name{} are tethered to reality,
  users must also supply \name{} with implementations (possibly
  blackbox) of these predicates. One possible implementation for
  $\I{hd}$ and $\I{mem}$ is: }

\begin{lstlisting}[language = ML, xleftmargin=20pt]
  let hd(s, u) =
     if Stack.is_empty s then false else (Stack.top s) == u

  let rec mem(s, u) =
     if Stack.is_empty s then false else (Stack.top s = u) ||
                                          mem(Stack.tail s, u)
\end{lstlisting}
\NV{\noindent where \lstinline{Stack.is\_empty}, \lstinline{Stack.top}
  and \lstinline{Stack.tail} refer to blackbox implementations of
  stack library methods.  }

\NV{While it is possible to na\"ively include a method predicate for
  each library method, such an approach may not be useful for
  verification. Some functions may be irrelevant to client assertions,
  unnecessarily increasing the set of possible specifications that
  must be considered. Conversely, the library may not include
  functions for desirable predicates; e.g., the $\S{Stack}$ library
  does not provide a \textsf{Stack.mem} function, although it is quite
  relevant for verifying our running example.}

\paragraph{Solution Space}

\begin{figure}
\begin{align*}
\begin{array}{ll}
\S{R_{is\_empty}} (\C{s}, \nu)  \mapsto  \forall u, (\nu \implies \neg \I{mem}(\C{s}, u)) \land (\neg \nu \land \I{hd}(\C{s}, u) \implies \I{mem}(\C{s}, u)) \\[0.5mm]
\S{R_{top}}(\C{s},\nu)  \mapsto  \forall u, \I{mem}(\C{s}, \nu) \land (\nu = u \iff \I{hd}(\C{s}, u))\\[0.5mm]
\S{R_{tail}} (\C{s}, \nu)  \mapsto \forall u, (\I{mem}(\C{s}, u) \implies (\I{mem}(\nu, u) \lor \I{hd}(\C{s}, u))) \land \\[0.5mm]
\qquad\qquad\qquad\ ((\I{mem}(\nu, u) \lor \I{hd}(\nu, u)) \implies \I{mem}(\C{s}, u))\\[0.5mm]
\S{R_{push}}(\C{x}, \C{s}, \nu)  \mapsto \forall u, (\I{mem}(\nu, u) \land \I{mem}(\C{s}, u) \implies \neg (\C{x} = u \lor \I{hd}(\nu, u))) \land \\[0.5mm]
\qquad\qquad\qquad\quad\ \ \ (\I{mem}(\nu, u) \land \neg \I{mem}(\C{s}, u) \implies (\C{x} = u \land \I{hd}(\nu, u)))\land \\[0.5mm]
\qquad\qquad\qquad\quad\ \ \ ((\C{x} = u \lor \I{hd}(\nu, u)
\lor \I{hd}(\C{s}, u) \lor \I{mem}(\C{s}, u)) \implies \I{mem}(\nu, u))
\end{array}
\end{align*}
\caption{\small Candidate verification interface for the $\S{concat}$
  example.}
\label{fig:consistent_solution}
\end{figure}

\NV{Our ultimate goal is to find a mapping from each placeholder
  predicate in $\textcolor{CBOrange}{\Sigma_{concat}}$ to an
  interpretation that entails
  $\textcolor{CBPurple}{\Phi_{\C{concat}}}$. We refer to such a
  mapping as a \emph{verification interface};
  \autoref{fig:consistent_solution} presents a potential verification
  interface for our running example.}  Not every mapping that ensures
the safety of $\C{concat}$ is reasonable, however. At one extreme,
interpreting every predicate as $\bot$ ensures the safety of the
client, but does not capture the behavior of any sensible stack
implementation.  Our goal, then, is to find interpretations that are
\emph{general} enough to cover a range of possible
implementations. From a purely logical perspective, this an instance
of a multi-abductive inference problem that tries to find the weakest
interpretations of $\S{R_{push}}$, $\S{R_{top}}$, $\S{R_{tail}}$ and
$\S{R_{is\_empty}}$ in terms of predicates \emph{mem} and \emph{hd}
such that the interpretations are self-consistent (i.e. $\bigwedge R_i \not \models \bot$) and which are sufficient
to prove the desired verification condition. While solutions to
the multi-abduction problem have been developed for domains that admit
quantifier elimination, e.g. linear integer arithmetic
constraints~\citet{AlbarghouthiMSS}, there is no purely logical
solution for formulae involving equalities over uninterpreted
functions.
An additional challenge in our setting is that we seek to
infer specifications consistent with the library's implementation, a
requirement that is absent in~\cite{AlbarghouthiMSS}.



\subsection{Data-Driven Abduction}



We overcome these challenges by adopting a data-driven approach to
abducing maximal library specifications, framing the problem as one of
training a Boolean classifier on a set of \emph{example behaviors} for
each function $f$. Under this interpretation, a classifier represents
a specification of the ``acceptable'' behaviors of $f$. Thus, the goal
of the specification inference stage of our algorithm is to learn a
set of classifiers that recognize the behaviors of each $f$ that a)
are consistent with $f$'s underlying implementation and b) preserve the
safety of the client program. As discussed at the start of this
section, this algorithm uses both an SMT-based verifier and a
property-based test generator as sources of training data for our
learner. The former identifies example behaviors consistent with $\neg
(\Sigma_\S{concat} \implies \Phi_\S{concat})$; these are labelled as ``negative'' examples,
so that the learned specifications can help the solver rule out
behaviors that are inconsistent with the semantics of the method
predicates or that produce unsafe executions (i.e.,
interpretations that would violate the postcondition). Example
behaviors drawn from tests are labelled as ``positive'', so that our learner is
biased towards explanations that are consistent with the (unknown)
implementations of library functions. Notably, our algorithm
generalizes this data-driven abduction procedure for individual
functions to the multi-abduction case, ensuring that discovered
interpretations are globally consistent over all library methods.

\begin{figure}[ht]
\includegraphics[width=300pt]{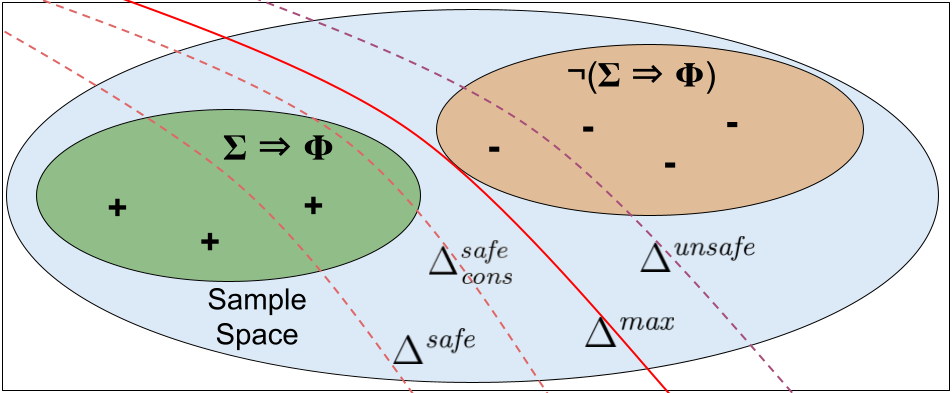}
\centering
\caption{\small Elements in the circle labelled $\Sigma \Rightarrow
  \Phi$ represent positive examples, while the orange $\neg(\Sigma
  \Rightarrow \Phi)$ circle contains negative examples. Our algorithm
  generates a mapping $\Delta$, which maps placeholder predicates to
  specifications, separating the two.}
\label{fig:safe_consistent_maximal}
\end{figure}

\autoref{fig:safe_consistent_maximal} depicts the space of example
behaviors for our learner, as well four potential verification
interfaces. Each of these represents a potential solution in the
\emph{hypothesis space} for this learner, which is tasked with
building a classifier that separates negative (-) and positive examples
(+) for each library function. The dashed purple line, labelled
$\Delta^\mathit{unsafe}$, represents an unsafe verification interface
that allows a client program to violate the desired postcondition. The
remaining red lines represent the range of safe verification
interfaces. The two dashed red lines represent the verification
interfaces that are sufficient to verify the client, but which are
suboptimal. $\Delta^\mathit{safe}$ is safe but inconsistent with the
observed behaviors of the library implementation, and is thus overly
restrictive.  \NV{$\Delta^\mathit{safe}_{\mathit{cons}}$} is safe and
consistent, but not maximal, as there exists a weaker verification
interface ($\Delta^\mathit{max}$) in the hypothesis space that is
still safe and consistent. \NV{Intuitively, the goal of our first
  phase is to identify $\Delta^\mathit{safe}_{\mathit{cons}}$, which
  is then weakened by the second phase to produce
  $\Delta^\mathit{max}$}.

\paragraph{Hypothesis space}
Our learner limits the shape of solutions it considers so that
inferred specifications are both amenable to automatic verification
and strong enough to verify specified postconditions. To enable
automated verification of client programs, potential specifications
are required to be prenex universally-quantified propositional
formulae over datatype values and variables representing arguments to
the predicates under consideration. Some possible specifications of
the library function $\S{push}$ in our running example include:
\[\S{push}(\I{x}, \I{l}) = \I{\nu}:
  \left\{\begin{array}{llllll}
           & \forall u, \I{mem}(l, u) \implies \I{mem}(\nu, u),
           & \forall u, \neg \I{mem}(\nu, u), \\
           & \forall u, u = x,
           & \forall u, u = x \iff \I{hd}(\nu, u), \\
           & \forall u, \I{mem}(l, u) \land \I{hd}(l,u),
           & \ldots
\end{array}\right\}\]
\noindent which contains, among other candidates, the desired specification.

All atomic literals in generated formulae are applications of
uninterpreted method predicates and equalities over quantified
variables, parameters, and return values of functions. The literals in
the above formulae are simply applications of \emph{hd} and \emph{mem}
to $l,\nu$ and $u$, and the equality $u=x$. We automatically discard
equalities between terms of different types, e.g. $l = x$.  The
\emph{feature set} for the predicate $\S{R_{push}}$, i.e. the set of
atomic elements used to construct its specification, is thus:
$\{\I{hd}(l, u),
\; \I{mem}(l, u),
\; \I{hd}(\S{\nu}', u),
\; \I{mem}(\S{\nu}, u),
\; x=u\}$.

\paragraph{Training Data} We now consider how to represent library behaviors
in a form that is useful for learning a solution in our hypothesis
space. To illustrate our chosen representation, consider the
counterexample produced by an off-the-shelf theorem prover when asked
to verify the formula
$\textcolor{CBOrange}{\Sigma_\C{concat}} \implies \textcolor{CBPurple}{\Phi_\C{concat}}$
from our running stack example, where the set of candidate
specifications are initialized to true (i.e., $\forall \S{f}.\, \S{R_f} \mapsto \top$):
\begingroup \setlength{\jot}{0pt}
\begin{align}\tag{\textsc{Cex}} \begin{split}
    \forall l\ u, \neg \I{hd}(l, u) \land \C{\nu_{top}} = \C{a} \land (\I{mem}(l, u) \iff ((l = \nu \lor l = \nu_{\S{concat}} \lor l = \nu_{\S{tail}}) \land u = \C{a}))
\end{split}
\end{align}
\endgroup


\noindent Intuitively, this counterexample asserts that the stacks
$\C{\nu}$, $\C{\nu_{tail}}$, and $\C{\nu_{concat}}$ contain exactly one
element, the constant $\C{a}$, and the other two stacks, $\C{s}_1$ and
$\C{s}_2$, are empty. This assertion indeed violates the second conjunct of
the postcondition,
$\I{mem}(\nu, u) \iff \I{mem}(\C{s}_1, u) \lor \I{mem}(\C{s}_2, u)$,
but it is inconsistent with the expected semantics of the library
functions, and can thus be safely ignored. The verifier generates this counterexample because the
interpretations of the placeholder predicates in
$\textcolor{CBOrange}{\Sigma_\C{concat}}$ are too permissive. In order
for the verifier to rule out this counterexample, the placeholder
predicates need to be strengthened to rule out this inconsistent
behavior.



Ignoring how we identify this counterexample as spurious for now, note
that there are many ways to strengthen these specifications. One
approach is to focus on one particular function at a time. For example,
we could choose to refine the specification of $\S{R_{top}}$ so that
it guarantees that $\nu_{\S{top}}$ is a member of
$\C{s}_1$. Alternatively, we could focus on $\S{R_{tail}}$, ensuring the
members of $\nu_{\S{tail}}$ are also contained by $\C{s}_1$. In
general, however, it may be necessary to strengthen multiple
specifications at once. Therefore, instead of focusing on one specification
at a time, we learn refined specifications simultaneously.

\begin{table}[]
\centering
\begin{tabular}{c|c|ccccc}
\toprule
$\S{R_{top}(s_1, \nu_{top})}$ & $\forall \S{u}$  &$\I{hd}(\S{s}_1, \S{u})$ &$\I{mem}(\C{s}_1, \S{u})$ &$\nu_{\S{top}} = \S{u}$ & &\\
\midrule
&$\S{u=\C{a}};$ & {\sf false} & {\sf false}  & {\sf true}  &  &\\
&$\S{u\not=\C{a}};$ & {\sf false} & {\sf false}  & {\sf false}  &  &\\
\bottomrule
$\S{R_{tail}(s_1, \nu_{tail})}$ & $\forall \S{u}$  &$\I{hd}(\C{s}_1, \S{u})$ &$\I{mem}(\C{s}_1, \S{u})$ &$\I{hd}(\nu_{\S{tail}}, \S{u})$ &$\I{mem}(\nu_{\S{tail}}, \S{u})$ &\\
\midrule
&$\S{u=\C{a}};$ & {\sf false} & {\sf false}  & {\sf false}  & {\sf true}  &\\
&$\S{u\not=\C{a}};$ & {\sf false} & {\sf false}  & {\sf false}  & {\sf false}  &\\
\bottomrule
$\S{R_{push}}(\nu_{top}, \nu_{concat}, \nu)$ &  $\forall \S{u}$ &$\I{hd}(\nu_{\S{concat}}, \S{u})$ &$\I{mem}(\nu_{\S{concat}}, \S{u})$ &$\I{hd}(\nu, \S{u})$ &$\I{mem}(\nu, \S{u})$ & $\nu_{\S{top}} = \S{u}$ \\
\midrule
&$\S{u=\C{a}};$ & {\sf false} & {\sf true}  & {\sf false}  & {\sf true} & {\sf true} \\
&$\S{u\not=\C{a}};$ & {\sf false} & {\sf false}  & {\sf false}  & {\sf false} & {\sf false} \\
\bottomrule
$\S{R_{is\_empty}(s_1, \nu_{is\_empty})}$ & $\forall \S{u}$ &$\I{hd}(\C{s}_1, \S{u})$ &$\I{mem}(\C{s}_1, \S{u})$ &$\nu_{\S{is\_empty}}$ & & \\
\midrule
&$\S{u=\C{a}};$ & {\sf false} & {\sf false}  & {\sf false}  & & \\
&$\S{u\not=\C{a}};$ & {\sf false} & {\sf false}  & {\sf false}  & & \\
\bottomrule
\end{tabular}
\caption{\small Potential negative feature vectors extracted from \sc{Cex}.
}
\label{tab:neg}
\end{table}


The first step to refining our placeholder specifications is to
extract data from \textsc{Cex} in a form that can be used to train a
classifier. We do this by using the assignments to the arguments of the
placeholder predicates in a counterexample to build \emph{feature
  vectors} that describe the valuations of method predicates and
equalities in the unsafe execution. \autoref{tab:neg} presents the
feature vectors extracted from \textsc{Cex}. The first column of this
table indicates the particular placeholder predicates that can be
strengthened to rule out this counterexample. The second column gives
the feature vectors for a particular instantiation of the quantified
variables ($e.g.,\ \C{u}$) of the placeholders. The subsequent columns
list applications of method predicates, with the rows underneath
listing the valuation of these predicates in the offending run. The
second row corresponds to the assertion that
$\neg \I{hd}(\C{s}_1, \C{a}) \land \neg \I{mem}(\C{s}_1, \C{a}) \land
\nu_{\S{top}} = \C{a}$, for example. A strengthening of the
specifications that disallows any one of these interpretations will
also rule out the corresponding unsafe run of the program. Put another
way, each row corresponds to a potential \emph{negative feature
vector}, and a \NV{classifier} (i.e., specification) for the
corresponding placeholder that disallows this feature will disallow
the counterexample.

The designation of these features as \emph{potentially} negative is
deliberate, as we only want to disallow features that are inconsistent
with the implementation of the library functions. \NV{As an example, the
  first feature vector for $\S{R_{top}}$ (the second row of
  Table~\ref{tab:neg}) states that the result of \lstinline|top s1| is
  not the head element of the input stack (since $\nu_{\S{top}} = \S{u}$ is
  true and $\I{hd}(\S{s}_1, \S{u})$ is false), and thus is
  inconsistent with \emph{any} reasonable implementation of $\S{top}$.
  In contrast, the next feature vector is compatible with an execution
  of $\S{top}$ where, e.g. $\S{top}([1])=1$ and $\S{u} = 2$.}
The second feature vector represents a behavior that is consistent with the
underlying library implementation and that should be allowed by the
learned specification. The consistency checker generates this positive
training data via random testing of the client program. To see how we
extract positive feature vectors from training data, consider the
execution of \lstinline|concat| with the inputs $\S{s}_1 = [\C{a}]$ and
$\S{s}_2 = [\C{b}]$, which produces the following assignment to program
variables: \begin{align*} \nu_{\S{is\_empty}} = \bot, \nu_{\S{top}} =
  \C{a}, \nu_{\S{tail}} = [], \nu_{\S{concat}} = [\C{b}], \nu =
  [\C{a;b}]
\end{align*}




Similar to how we built negative feature vectors from
\textsc{Cex}, we can construct feature vectors for each function
specification from these assignments. \autoref{tab:pos} illustrates
the feature vectors corresponding to this assignment. Consider the
second row where $\S{u}$ is instantiated with $\C{a}$: under this
assignment, $\I{hd}(\S{s}_1, \S{u}) \equiv \S{hd}([\C{a}], \C{a})$ is
true, as are $\I{mem}(\C{s}_1, \S{u}) \equiv \S{mem}([\C{a}], \C{a})$
and $\nu_{\S{top}} = \S{u} \equiv \C{a} = \C{a}$.

\begin{table}[]
\centering
\begin{tabular}{c|c|ccccc}
\toprule
$\S{R_{top}(s_1, \nu_{top})}$ &$\forall \S{u}$ &$\I{hd}(\S{s}_1, \S{u})$ &$\I{mem}(\C{s}_1, \S{u})$ &$\nu_{\S{top}} = \S{u}$ & &\\
\midrule
\multirow{2}{*}{}&$\S{u=\C{a}};$ & {\sf true} & {\sf true}  & {\sf true}  &  &\\
&$\S{u=\C{b}};$ & {\sf false} & {\sf false}  & {\sf false}  &  &\\
\bottomrule
$\S{R_{tail}(s_1, \nu_{tail})}$ & $\forall \S{u}$ &$\I{hd}(\C{s}_1, \S{u})$ &$\I{mem}(\C{s}_1, \S{u})$ &$\I{hd}(\nu_{\S{tail}}, \S{u})$ &$\I{mem}(\nu_{\S{tail}}, \S{u})$ &\\
\midrule
\multirow{2}{*}{}&$\S{u=\C{a}};$ & {\sf true} & {\sf true}  & {\sf false}  & {\sf false}  &\\
&$\S{u=\C{b}};$ & {\sf false} & {\sf false}  & {\sf false}  & {\sf false}  &\\
\bottomrule
$\S{R_{push}}(\nu_{top}, \nu_{concat}, \nu)$ & $\forall \S{u}$ &$\I{hd}(\nu_{\S{concat}}, \S{u})$ &$\I{mem}(\nu_{\S{concat}}, \S{u})$ &$\I{hd}(\nu, \S{u})$ &$\I{mem}(\nu, \S{u})$ & $\nu_{\S{top}} = \S{u}$ \\
\midrule
\multirow{2}{*}{}&$\S{u=\C{a}};$ & {\sf false} & {\sf false}  & {\sf true}  & {\sf true} & {\sf true} \\
&$\S{u=\C{b}};$ & {\sf true} & {\sf true}  & {\sf false}  & {\sf true} & {\sf false} \\
\bottomrule
$\S{R_{is\_empty}(s_1, \nu_{is\_empty})}$ & $\forall \S{u}$ &$\I{hd}(\C{s}_1, \S{u})$ &$\I{mem}(\C{s}_1, \S{u})$ &$\nu_{\S{is\_empty}}$ & & \\
\midrule
\multirow{2}{*}{}&$\S{u=\C{a}};$ & {\sf true} & {\sf true}  & {\sf false}  & & \\
&$\S{u=\C{b}};$ & {\sf false} & {\sf false}  & {\sf false}  & & \\
\bottomrule
\end{tabular}
\caption{\small Positive feature vectors extracted from executing
  \lstinline|concat([a], [b])|.
}
\label{tab:pos}
\end{table}

\begin{table}[]
\renewcommand{\arraystretch}{0.6}
\begin{tabular}{c|cccccc}
\toprule
$\S{R_{top}(s_1, \nu_{top})}$ &$\I{hd}(\S{s}_1, \S{u})$ &$\I{mem}(\C{s}_1, \S{u})$ &$\nu_{\S{top}} = \S{u}$ & &\\
\midrule
- & {\sf false} & {\sf false}  & {\sf true}  &  &\\
\midrule
+ & {\sf true} & {\sf true}  & {\sf true}  &  &\\
+ & {\sf false} & {\sf false}  & {\sf false}  &  &\\
\bottomrule
\end{tabular}
\caption{\small Disjoint positive and negative feature vectors of $\S{R_{top}}$.}
\label{tab:pos-neg}
\end{table}

\paragraph{Classification} In order to train a classifier, we
need to label the extracted feature vectors as either positive or
negative. In other words, we need to identify behaviors that should (and should
not) be allowed by the inferred specification. Assigning labels is not
as straightforward as labelling the feature vectors extracted from
counterexamples as negative and those extracted from testing as
positive, as the two sets can overlap. We observe this in the
vectors of $\S{R_{top}}$ from \autoref{tab:neg} and \autoref{tab:pos}:
the interpretation $\I{hd}(\S{s}_1, \S{u}) \mapsto \S{false};
\I{mem}(\C{s}_1, \S{u}) \mapsto \S{false}; \nu_{\S{top}} = \S{u}
\mapsto \S{false};$ occurs in both tables. Intuitively,
we do not want to strengthen the specification of $\S{R_{top}}$ to
rule out this interpretation, as the positive sample is a witness that
this execution is consistent with the implementation of $\S{R_{top}}$.
\NV{Ultimately, therefore, the specification must be relaxed to allow
the execution. In cases where a negative feature vector conflicts with
a positive feature vector, we identify the potential negative feature vector as positive
and remove it from the learner's negative feature vector set.}
This
strategy is similar to the one used by ~\citet{RepInvGen} to deal with
inductiveness counterexamples.

Thankfully, as long as the counterexample set contains at least
one feature vector not known to be consistent with the underlying
library implementation, we can strengthen the specifications to
disallow it. The first feature vector for $\S{R_{top}}$ in \autoref{tab:neg}
represents one such infeasible execution. This vector encodes the case
where $u$ is the output of $\S{top}$ but is not a member or head of the
input stack. Clearly, no reasonable implementation would support such
an interpretation.
We use this observation to label as ``negative'' those feature vectors
that are extracted from a counterexample but do not appear in the set
drawn from a concrete execution. \autoref{tab:pos-neg} shows the
partition of positive and negative feature vectors for $\S{R_{top}}$.

Given this labelled set of positive and negative feature vectors,
\NV{our data-driven learner} builds a separator over the training
data. One such classifier (formula) for the data in
\autoref{tab:pos-neg} is
$\nu_{\S{top}} = \S{u} \implies \I{hd}(\S{s}_1, \S{u})$. Substituting
similarly learned specifications for the other library functions in
$\textcolor{CBOrange}{\Sigma_\C{concat}}$ equips the SMT solver with enough constraints to
rule out \textsc{Cex} \NV{while maintaining the invariant that the
  learned specifications are also consistent with the underlying
  library implementations}. Additional iterations of this
counterexample-guided refinement loop gather additional positive and
negative features, eventually producing the specifications for the
library functions presented in \autoref{fig:consistent_solution}.


\paragraph{Identifying Spurious Counterexamples}
\NV{Thus far, we have
only considered spurious counterexamples generated by the safety
checker. Counterexamples, however, can also result from an
incorrect client assertion. For example, suppose the client
(unsoundly) asserts}:
\begin{align*}
    \forall u, \I{mem}(\nu, u) \implies \I{mem}(\C{s}_2, u)
\end{align*}
This assertion is wrong, assuming reasonable implementations of $\S{top}$ and
$\S{push}$, as the elements of the result stack $\nu$ can also come from
$\C{s}_1$. \NV{We distinguish counterexamples corresponding to actual
  safety violations by first checking if all the feature
  vectors extracted from the counterexample are included in the set of
  known positive feature vectors. For example, given this unsound assertion,}
the verifier may produce the following counterexample:

\begingroup \setlength{\jot}{0pt}
\begin{align}\tag{\textsc{Cex'}} \begin{split}
    \forall l\ u,& (\I{hd}(l, u) \iff ((l = \nu \lor l = \C{s}_1) \land u = \C{a})) \land \C{\nu_{top}} = \C{a} \land \\
    &(\I{mem}(l, u) \iff ((l = \nu \lor l = \C{s}_1) \land u = \C{a}))
\end{split}
\end{align}

\begin{table}[]
\centering
\begin{tabular}{c|c|ccccc}
\toprule
$\S{R_{top}(s_1, \nu_{top})}$ &$\forall \S{u}$ &$\I{hd}(\S{s}_1, \S{u})$ &$\I{mem}(\C{s}_1, \S{u})$ &$\nu_{\S{top}} = \S{u}$ & &\\
\midrule
&$\S{u=\C{a}};$ & {\sf true} & {\sf true}  & {\sf true} \\
&$\S{u\not=\C{a}};$ & {\sf false} & {\sf false}  & {\sf false} \\
\bottomrule
\end{tabular}
\caption{\small Feature vectors for $\S{R_{top}}$ extracted from \sc{Cex'}.}
\label{tab:neg2}
\end{table}
\endgroup

\NV{\autoref{tab:neg2} shows all of the feature vectors for
  $\S{R_{top}}$ that are extracted from this counterexample}. Since these are a subset of
the positive feature vectors from \autoref{tab:pos}, there
are no feature vectors that can be labeled as negative, and there are
thus no new bad behaviors that the learner can use to generate a
refined specification mapping that rejects the counterexample.
In this scenario, our algorithm tries to
discover concrete values of $\C{s}_1$ and $\C{s}_2$ consistent with
\textsc{Cex'}; that is, $\C{s}_1$ only contains $\C{a}$ and $\C{s}_2$
is empty. When called with these parameters, $\S{concat}([\C{a}], [])$
will return $\nu \equiv [\C{a}]$, which \name{} returns as a witness
of an unsafe execution.

\NV{Note that this situation may also occur when the feature set is
  not large enough, as the specifications in the corresponding
  hypothesis space are not expressive enough to identify a spurious
  counterexample.  Thus, if we are not able to find inputs that
  trigger a safety violation, we grow the feature set by increasing
  the number of quantified variables (\emph{e.g.} from $u$ to $u,v$)
  so that \name{} can explore a richer space of specifications.}

\paragraph{Weakening.}
\NV{While the above strategy is guaranteed to find a safe and
  consistent verification interface when one exists, the solutions it produces may
  still be suboptimal, as illustrated in \autoref{fig:safe_consistent_maximal}.
  For example, the first conjunct of the specification for
  \lstinline|push| in \autoref{fig:consistent_solution} states that any
  existing member of both the input and output stacks should not be
  the same as the element being added to the stack; that is,
  \lstinline|push| always produces a stack with no duplicates.  This
  specification is too restrictive, however, as it disallows
  reasonable behaviors such as \lstinline|push([1;2], 1)| = \lstinline|[1;1;2]|.
  However, if our sampler never generates an observation corresponding
  to this behavior, e.g.  $x \equiv 1$, $\C{s} \equiv [1;2]$ and
  $\nu_{push} \equiv [1;1;2]$, the candidate specification for
  \lstinline|top| produced by \name{}'s first phase will incorrectly
  disallow it. }

In other words, our reliance on testing to identify and label negative feature vectors
may result in initial specifications that are overfitted to the
examples enumerated by the test generator. There are two potential
reasons such a positive example might be missed: (1) the input space of
the program might be too large for a test generator to effectively
explore, and (2) the provided implementation may simply not exhibit
this behavior (e.g., it may be the case that the implementation of
\lstinline|push| that we are trying to verify against does indeed remove
duplicates).  While exhaustive or more effective enumeration can
address the first cause, it cannot remedy the second. \name{}'s
weakening phase helps ameliorate both issues.


Our weakening algorithm iteratively weakens candidate specifications,
focusing on one library function at a time. To weaken the
specification of \lstinline|push|, for example, we fix the specifications of
the other library functions to their assignments in the current
verification interface, and then try to find a maximal weakening of
$\S{R_{push}}$ that admits a larger set of implementations of
\lstinline|push|. \NV{To do so, \name{} attempts to discover additional \emph{weakening
    feature vectors} for $\S{R_{push}}$, or feature vectors
  corresponding to behaviors disallowed by the current specification
  but which would not lead to a violation of client safety. One
  possible weakening feature vector for our current example is shown
  in \autoref{tab:weaken}. Here, the head of both $\S{s}_1$ and the
  result of the recursive call is $\S{1}$; this scenario
  is mistakenly disallowed by the specification of \lstinline|push| in
  \autoref{fig:consistent_solution}. }



\begin{table}[]
\centering
\begin{tabular}{c|c|ccccc}
\toprule
$\S{R_{push}}(\nu_\S{top}, \nu_\S{concat}, \nu)$ &  $\forall \S{u}$ &$\I{hd}(\nu_{\S{concat}}, \S{u})$ &$\I{mem}(\nu_{\S{concat}}, \S{u})$ &$\I{hd}(\nu, \S{u})$ &$\I{mem}(\nu, \S{u})$ & $\nu_{\S{top}} = \S{u}$ \\
\midrule
&$\S{u=\C{1}};$ & {\sf true} & {\sf true}  & {\sf true}  & {\sf true} & {\sf true} \\
\bottomrule
\end{tabular}
\caption{\small A weakening feature vector of $\S{R_{push}}$ in \autoref{fig:consistent_solution}}
\label{tab:weaken}
\end{table}

\NV{\name{} repeatedly queries the verifier to identify weakening
  feature vectors for \lstinline|push|, indicating that the current
  specification is maximal when none can be found. It then moves on to
  the next function specification, iteratively weakening each until a
  fixpoint is reached. } \autoref{fig:maximal_solution} shows the
\NV{maximal verification interface \name{} builds by weakening the candidate specifications in
  \autoref{fig:consistent_solution}}. Compared with
\autoref{fig:consistent_solution}, the specification of \lstinline|push| now permits duplicate
elements in stacks. In addition, the specification of
\lstinline|is_empty| has been simplified by removing the redundant conjunction
$\neg \nu \land \I{hd}(\C{s}, u) \implies \I{mem}(\C{s}, u)$, as
$\I{hd}(\C{s}, u) \implies \I{mem}(\C{s}, u)$ can never be violated by
a concrete stack value thanks to the observed semantics of $\I{hd}$
and $\I{mem}$.

\begin{figure}[ht!]
\begin{align*}
\begin{array}{ll}
\S{R_{is\_empty}} (\C{s}, \nu)  \mapsto  \forall u, (\nu \implies \neg \I{mem}(\C{s}, u)) \\[0.5mm]
\S{R_{top}}(\C{s},\nu)  \mapsto  \forall u, \nu = u \iff (\I{hd}(\C{s}, u) \land \I{mem}(\C{s}, u)) \\[0.5mm]
\S{R_{tail}} (\C{s}, \nu)  \mapsto \forall u, (\I{mem}(\C{s}, u) \implies (\I{mem}(\nu, u) \lor \I{hd}(\C{s}, u))) \land \\[0.5mm]
\qquad\qquad\qquad\ ((\I{mem}(\nu, u) \lor (\I{hd}(\nu, u) \land \I{hd}(\C{s}, u))) \implies \I{mem}(\C{s}, u))\\[0.5mm]
\S{R_{push}}(\C{x}, \C{s}, \nu)  \mapsto \forall u, (\I{mem}(\nu, u) \land \neg \C{x} = u \implies \I{mem}(\C{s}, u) \land\neg  \I{hd}(\nu, u)) \land \\[0.5mm]
\qquad\qquad\qquad\quad\ \ \ (\C{x} = u \lor \I{mem}(\C{s}, u) \implies \I{mem}(\nu, u)) \land (\neg \I{mem}(\nu, u) \land \I{hd}(\nu, u) \implies \I{hd}(\C{s}, u))
\end{array}
\end{align*}
\caption{\small Maximal specifications for \lstinline|Stack| operations. The special variable $\nu$ represents
the result of the method.}
\label{fig:maximal_solution}
\end{figure}

\section{Problem Formulation}
\label{sec:formalization}
Having completed a high-level tour of \name{} in action, we now
present a precise description of the specification synthesis problem
and our data-driven inference procedure. We consider functional
programs that use data structure libraries providing functions to
access and construct instances of inductively-defined algebraic
datatypes (e.g., list, stacks, trees, heaps, tries, etc.).

In the remainder of the paper, we use $(\Sigma, \Phi)$ to refer to the
\emph{verification query} whose validity we are attempting to
establish. These structures serve the same role as verification
conditions in a typical verification framework. The first component of
this query, $\Sigma$, is a conjunction of applications of
specification placeholders ($R_f$) to arguments; these represent the
library method calls made by the client program. The second component,
$\Phi$, represents the client program's pre- and post-conditions,
encoded as sentences built from logical connectives ($\wedge$, $\vee$,
$\implies$) over prenex universally-quantified propositional
formulae. \NV{Each verification query corresponds to a control flow
  path in the client program; the full algorithm considers the
  conjunction of all these verification queries at once. To keep the
  formalization and the description of our algorithms concise, our
  description considers a single verification query in isolation. The
  extension to sets of verification queries is provided \releasesource{}.
  }

\begin{definition}[Problem Definition]
A given verification query $(\Sigma,\Phi)$ with unknown library
functions $F$ has the form $\Sigma \implies \Phi$, where:
\begin{align*}
    &\Sigma \equiv (\bigwedge_{f \in F, i = 0}^n R_f(\vec{x_{i}})
      \land \bigwedge_{x, x' \in \cup_{0\le j \le i} \vec{x_j} \cup \mathcal{C}} x = x')
\end{align*}
Here, the equality constraints are either between program variables
($x,x'$) or between variables and constants $\mathcal{C}$ of some base type
(e.g., Booleans and integers).  Each $R_f(\vec{x_{i}})$ in $\Sigma$ is an
application of a placeholder predicate to some arguments; the
conjunction of these placeholder applications
and equality constraints represents a sequence of library method
invocations in one control-flow path of the client.
\end{definition}
To model the input and output behaviors of the blackbox
implementations of library functions and method predicates, our
formalization relies on a pair of partial functions with the same
signature as the implementations. We use partial functions to reflect
the fact that we can only observe a subset of the full behaviors of
these implementations when searching for specifications.
\begin{definition}[Specification Configuration]
  Let $P$ be a set of method predicates and $F$ be the set of functions in a
  library used by the client.  Let $\Gamma_{f}$ be a partial function
  from the domain of  $f \in F$ to its codomain, and
  $\Gamma_{p}$ be a partial function with the same signature as
  $p \in P$. Let $\Gamma_P = \bigcup_{p \in P} \Gamma_p$ and
  $\Gamma_F = \bigcup_{f \in F} \Gamma_f$.  A \emph{specification
    configuration} is a 5-tuple
  $((\Sigma, \Phi), P, F, \Gamma_{P}, \Gamma_{F})$, where
  $(\Sigma, \Phi)$ is the verification query extracted from the client.
\end{definition}

\textbf{Example.} The specification configuration of our running
example consists of a verification query
$(\textcolor{CBOrange}{\Sigma_{concat}},
\textcolor{CBPurple}{\Phi_{\C{concat}}})$, a method predicate set
$\{\I{mem}\I{hd}\}$, and library functions
$\{\S{push}$, $\S{is\_empty}$, $\S{top}$, $\S{tail}\}$. The partial
functions in $\Gamma_F$ and $\Gamma_P$ abstract over observations on
their corresponding blackbox implementations; for an execution that
produces the feature vectors shown in \autoref{tab:pos}, they are:

\begin{align*}
    \Gamma_{\I{mem}} ~\equiv~ & \lambda (l \in \{\S{s_1}, \S{s_2}, \nu_{\S{tail}}, \nu_{\S{concat}}, \nu\}),u.\\
    &(u = \C{a} \land (l = \S{s_1} \lor l = \nu)) \lor (u = \C{b} \land (l = \S{s_2} \lor l = \nu_{\S{concat}} \lor l = \nu)) \\
    \Gamma_{\I{hd}} ~\equiv~ & \lambda (l \in \{\S{s_1}, \S{s_2}, \nu_{\S{tail}}, \nu_{\S{concat}}, \nu\}),u.\\
    &(u = \C{a} \land (l = \S{s_1} \lor l = \nu)) \lor (u = \C{b} \land (l = \S{s_2} \lor l = \nu_{\S{concat}})) \\
    \Gamma_{\S{push}} ~\equiv~& \lambda x, (l \in \{\nu_{\S{concat}})\}). \nu \\
    \Gamma_{\S{is\_empty}} ~\equiv~& \lambda (l \in \{\S{s_1}\}), \mathit{false} \\
    \Gamma_{\S{top}} ~\equiv~& \lambda (l \in \{\S{s_1}\}), \C{a} \\
    \Gamma_{\S{tail}} ~\equiv~& \lambda (l \in \{\S{s_1}\}),  \nu_{\S{tail}}
\end{align*}
where stack arguments are limited to values in this particular
execution, e.g. $\S{s_1}$ or $\nu_{\S{tail}}$.


Given a specification configuration as input, the output of our
verification pipeline is a \emph{verification interface} ($\Delta$), a
logical interpretation of the method predicates that maps each
placeholder predicate for a function $f \in F$ to a
universally-quantified propositional formula over the parameters and
result of $f$.  We impose two requirements on $\Delta$.  The first is
\emph{safety}: an underlying theorem prover (e.g., a SMT solver) must
be able to prove $\Sigma[\Delta] \implies \Phi$, where
$\Sigma[\Delta]$ denotes the formula constructed by replacing all
occurrences of specification placeholders with their interpretation in
$\Delta$, and $\Sigma \implies \Phi$ is the verification query built
from a client program:
\begin{definition}[Safe Verification Interface]
  For a given verification query $(\Sigma, \Phi)$, a verification
  interface $\Delta$ is \emph{safe} when:
\begin{enumerate}
    \item it makes the VC valid: $\Sigma[\Delta] \models \Phi$, and
    \item is not trivial: $\Sigma[\Delta] \not\models \bot$
\end{enumerate}
\label{def:safe}
\end{definition}
\noindent In addition to safety, we also desire that any proposed mapping
$\Delta$ be \emph{consistent} with the provided implementations of
method predicates and library functions, \emph{i.e.} that $\Delta$ must
accurately represent their observed behavior. Formally:
\begin{definition}[Interface Consistency]
  A verification interface $\Delta$ is consistent with
  $\Gamma_{P}$ and $\Gamma_{F}$ when all specifications in
  $\Delta$ are consistent with the inputs on which $\Gamma_P$ and $\Gamma_F$ are
  defined.  Formally,
\begin{align*}
  \forall f \in F, \Gamma_f \in \Gamma_F, \Gamma_f(\vec{\alpha}) = \nu \Rightarrow
     \Delta(R_f)(\vec{\alpha},\nu)[\Gamma_P]
\end{align*}
The expression $\Delta(R_f)(\vec{\alpha}, \nu)$ denotes the instantiation of the
formula bound to $R_f$ in $\Delta$ with the input arguments
$\vec{\alpha}$ and observed output $\nu$.  The expression
$\Delta(R_f)(\vec{\alpha},\nu)[\Gamma_P]$ replaces all free
occurrences of $p$ in $\Delta(R_f)(\vec{\alpha},\nu)$ with
$\Gamma_p$ where $\Gamma_p \in \Gamma_P$.
\label{def:mapconsistency}
\end{definition}
This definition thus relates the observed behavior of a library method
on test data, encoded by $\Gamma_P$ and $\Gamma_F$ with its logical
characterization provided by $\Delta$.  Note that there may be many
possible verification interfaces for a given specification
configuration. In order to identify the best such interface, we use an
ordering based on a natural logical inclusion property:
\begin{definition}[Interface Order $\succ$]
The verification interface ${\Delta'}$ is weaker ($\succ$) than $\Delta$ when,
\begin{enumerate}
\item The two interfaces contain the same functions:
  $dom(\Delta) = dom(\Delta')$
\item They are not equal:
  $\exists R_f \in dom(\Delta), \Delta(R_f) \centernot\iff \Delta'(R_f)$
    \item The specifications in $\Delta'$ are logically weaker that those in $\Delta$: $\forall R_f \in dom(\Delta), \Delta(R_f) \implies \Delta'(R_f)$
\end{enumerate}
\label{def:order}
\end{definition}
\noindent Intuitively, weaker verification interfaces are preferable
because they place fewer restrictions on the behavior of the
underlying implementation. Given an ordering over verification
interfaces, we seek to find the weakest safe and consistent interface,
\emph{i.e.} one that imposes the fewest constraints while still enabling
verification of the client program.
\begin{definition}[Maximal Verification Interface]
  For a specification configuration $((\Sigma, \Phi)$, $P$, $F$,
  $\Gamma_{P}$, $\Gamma_{F})$, $\Delta$ is a maximal verification interface when:
\begin{enumerate}
    \item $\Delta$ is safe for the verification query $(\Sigma, \Phi)$.
    \item $\Delta$ is consistent with $\Gamma_{P}$ and $\Gamma_{F}$.
    \item For a given bound on the number of quantified variables $k$
      used by the specifications in $\Delta$, there is no safe and consistent
      interface $\Delta'$ whose specifications use at most $k$ quantified variables such that
      $\Delta' \succ \Delta$.
\end{enumerate}
\label{def:max_saf_consistent}
\end{definition}
We now refine our expectation for the output of our verification
pipeline to be not just any safe and consistent verification
interface, but also a \emph{maximal} one.  Notice that our notion of
maximality is parameterized by the number of quantified variables used
in the interpretation.  As this bound increases, we can always find a
weaker specification mapping. Thus we frame our definition of
maximality to be relative to the number of quantified variables in the
specification.

\section{Learning Library Specifications}
\label{sec:learning}
As \autoref{sec:motivation} outlined, \name{} frames the search for a
safe verification interface as a data-driven learning problem. At a high level,
the goal of learning is to build a \emph{classifier} (a function from unlabeled
data to a label) from a set of labeled data. More precisely, our goal is to
learn classifiers for each of the library functions in a specification
configuration that can correctly identify any input and output behavior that
could induce an unsafe execution in the client.

Our first challenge is to find an encoding of program executions that
is amenable to a data-driven learning framework. To begin, we need to
identify the salient \emph{features} used by a classifier to make its
decisions.

\begin{definition}[Feature]
  A \emph{feature} of a function for a set of variables
  $\vec{x}$ is a method predicate applied to elements of $\vec{x}$ or
  equalities between variables in $\vec{x}$.
\end{definition}
\noindent A feature is similar to a literal in first-order logic, but
does not allow for method predicates as arguments (e.g.
$\I{hd}(l, \I{mem}(l, u))$) or constant arguments (e.g.
$\I{hd}(l, \C{3})$).

\begin{definition}[Feature Set]
  \label{def:featureset}
  The \emph{feature set} of a function $f$ with method predicates $P$
  and quantified variables $\vec{u}$, denoted as
  $\mathcal{S} \equiv \mathit{FSet}(P, f(\vec{\alpha}_f) = \nu_f,
  \vec{u})$, is a list of all well-typed features in $P$ for the set
  of variables $\vec{\alpha}_f \cup \{ \nu_f \} \cup \vec{u}$ which is
  \emph{minimally linearly independent}:
  \begin{align*}
      \forall \eta' \not \in \mathcal{S}, \exists \vec{\eta} \subseteq \mathcal{S}, \eta' \iff \bigwedge \vec{\eta}
  \end{align*}
\end{definition}

\textbf{Example.} The feature set for the function
$\S{top} : list_a \rightarrow a$ from the \texttt{Stack}
library, where $a$ is some base type, for predicate set
$P \equiv \{\I{hd},\I{mem}\}$, equality operation $\I{=}_{a}$ and quantified
variables $\vec{u} \equiv \{ u:a \}$ is
$\mathit{FSet}(P, \S{top}(l) = \nu, \vec{u}) \equiv
[\I{hd}(l,u),\; \I{mem}(l,u),\; \nu \I{=}_{a} u]$. Note that the features
$\I{mem}(\nu,u)$ and $l =_a u$ are not included in this set because
they are not well-typed. The feature $\I{hd}(l,\nu)$, on the other
hand, is omitted because it can be represented by
$\I{hd}(l,u) \land \nu \I{=}_{a} u$ and is thus not linearly
independent with respect to the other features in the set.  We use \emph{feature
  vectors} to encode the features of observed tests:
\begin{definition}[Feature Vector]
\label{def:featurevectors}
A feature vector is a vector of Booleans that represents
the value of each feature in the feature set for some test.
\end{definition}

We also need to define the \emph{hypothesis space} of possible
solutions considered by our learning system. To easily integrate
learned classifiers into the underlying theorem prover, we choose to
represent such solutions as Boolean combinations over terms consisting
of applications of interpreted base relations and uninterpreted
functions. In order to preserve decidability, we limit this space to a
subset of effectively propositional sentences. \NV{This limitation was
expressive enough for all of our benchmarks.\footnote{\NV{The main sorts of
properties that we do not support as a consequence of this choice
are those which use quantifier alternation, e.g., for every element in
a stream, there exists another larger element that appears after
it: ($\forall u, \exists v, \mathit{mem}(l,u) \implies (\mathit{mem}(l,v) \land
\mathit{ord}(l,u,v) \land u \le v)$).}}
}


\begin{definition}[Hypothesis Space]
  The \emph{hypothesis space} of specifications for a library function
  $f$, method predicate set $P$, and quantified variables $\vec{u}$ is
  the set of formulas in prenex normal form with the quantifier prefix
  $\forall \vec{u}$, and whose bodies are built from
  $FSet(P, f(\vec{\alpha}) = \nu, \vec{u})$, the logical connectives
  $\{\land,\lor,\neg, \implies\}$, and Boolean constants $\top$ ({\sc
    true}) and $\bot$ ({\sc false}). The hypothesis space of $f$ over
  $P$ and $\vec{u}$ is denoted
  $Hyp(P, f(\vec{\alpha}) = \nu, \vec{u})$.
\end{definition}

In order to classify feature vectors, we ascribe them a semantics in
logic:
\begin{definition}[Unitary classifier]
  For a given feature vector $\mathit{fv}$ in a feature set $\mathcal{S} \equiv \mathit{FSet}(P, f(\alpha) = \nu, \vec{u})$, the
  \emph{logical embedding} of $\mathit{fv}$ is a formula encoding the
  assignment to its features:
  \begin{align*}
    \unitaryc{\I{fv}} \equiv \forall \vec{u}, \bigwedge_{i = 0}^{|\mathcal{S}|} \mathcal{S}[i] \iff \mathit{\I{fv}}[i]
  \end{align*}
  We say that a classifier $\phi$ labels a feature vector $\I{fv}$ as
  positive when $\unitaryc{\I{fv}} \implies \phi$, and negative
  otherwise.
\end{definition}
\textbf{Example.} Given the classifier
$\phi \equiv \forall u, \I{hd}(\S{s}_1, \S{u}) \implies \nu_{\S{top}}
= \S{u}$ for the $\S{top}$ function from \autoref{sec:motivation},
\NV{the first row in \autoref{tab:pos-neg} corresponds to the feature vector $\mathit{\I{fv}}^- \equiv \{\I{hd}(\S{s}_1, \S{u}) \mapsto \sf false;
\I{mem}(\C{s}_1, \S{u}) \mapsto \sf false; \nu_{\S{top}} = \S{u}
\mapsto \sf true\}$. The unitary
classifier for $\mathit{\I{fv}}$ is $\unitaryc{\I{fv}^-} \equiv
\forall u, \neg \I{hd}(\S{s}_1, \S{u}) \land \neg \I{mem}(\C{s}_1,
\S{u}) \land \nu_{\S{top}} = \S{u}$. $\I{fv}$ is labelled negative by
$\phi$, as $\unitaryc{\I{fv}^-} \centernot\implies \phi$.
The other two feature vectors in \autoref{tab:pos-neg} are labeled as
positive by $\phi$.}

\begin{definition}[Classification]
  For a given classifier $\phi$ and feature set $\mathcal{S}$, it is
  straightforward to partition the feature vectors of $\mathcal{S}$
  into positive ($\phi^+$) and negative ($\phi^-$) sets:
  \begin{align*}
    \phi^+ \equiv\quad & \{\mathit{fv} \in 2^\mathcal{S}~|~\unitaryc{\mathit{fv}} \implies \phi\} \\
    \phi^- \equiv\quad & \{\mathit{fv} \in 2^\mathcal{S}~|~\unitaryc{\mathit{fv}} \centernot\implies \phi\}
  \end{align*}
  Notice that these two sets are trivially disjoint:
  $\phi^+ \cap \phi^- = \emptyset$.
  \label{def:classification}
\end{definition}

For a particular configuration, we can straightforwardly lift this
partitioning to verification interfaces:
\begin{align*}
  \Delta(R_f)(\vec{\alpha}, \nu)^+ \equiv\quad & \{\mathit{fv} \in 2^{\mathit{FSet}(P, f(\vec{\alpha}) = \nu, \vec{u})}~|~\unitaryc{\mathit{fv}} \implies \Delta(R_f)(\vec{\alpha}, \nu)\} \\
  \Delta(R_f)(\vec{\alpha}, \nu)^- \equiv\quad & \{\mathit{fv} \in  2^{\mathit{FSet}(P, f(\vec{\alpha}) = \nu, \vec{u})}~|~\unitaryc{\mathit{fv}} \centernot\implies \Delta(R_f)(\vec{\alpha}, \nu)\}
\end{align*}




\subsection{Learning Safe and Consistent Verification Interfaces}
We now confront the challenge of how to generate training data from a
specification configuration in a way that guarantees the safety of the
learned formulas (classifiers). To do so, we extract feature
vectors from a set of \emph{logical samples}:
\begin{definition}[Sample]
\label{def:sample}
A \emph{sample} $s$ of a formula $\phi$ is an instantiation of
its quantified variables and a Boolean-valued interpretation for each
application of a method predicate to those variables in $\phi$, which
we denote as $s \models \phi$. The positive and negative samples of a
verification query $(\Sigma,\Phi)$ are samples of $\Phi$ and
$\neg \Phi$, respectively.
\end{definition}
Intuitively, the positive samples of a verification query
correspond to safe executions of a client program, while negative
samples represent potential violations that safe verification
interfaces need to prevent. For example, \textsc{Cex} from
\autoref{sec:motivation} corresponds to the following
negative sample of $\textcolor{CBPurple}{\Phi_{\C{concat}}}$\NV{\footnote{\NV{The interpretation of method predicates are represented as binary relations.}}}:
\begingroup \setlength{\jot}{0pt}
\begin{align*}\tag{$s^-$}
\{&\C{s}_1 \mapsto \C{l}_0; \C{s}_2 \mapsto \C{l}_1; \nu \mapsto \C{l}_2; \nu_{\S{top}} \mapsto \C{a}; \nu_{\S{tail}} \mapsto \C{l}_3; \nu_{\S{concat}} \mapsto \C{l}_4; \nu_{\S{is\_empty}} \mapsto \bot; \\
&\I{hd}(l, u)\equiv \{\}; \I{mem}(l, u) \equiv \{(\C{l}_2, \C{a}); (\C{l}_3, \C{a}); (\C{l}_4, \C{a})\}\}
\end{align*}
\endgroup
and the following sample, extracted from a concrete input and client
execution result, is positive: \begingroup \setlength{\jot}{0pt}
\begin{align*}\tag{$s^+$}
\{&\C{s}_1 \mapsto \C{l}_0; \C{s}_2 \mapsto \C{l}_1; \nu \mapsto \C{l}_2; \nu_{\S{top}} \mapsto \C{a}; \nu_{\S{tail}} \mapsto \C{l}_3; \nu_{\S{concat}} \mapsto \C{l}_4; \nu_{\S{is\_empty}} \mapsto \bot; \\
&\I{hd}(l, u)\equiv \{(\C{l}_0, \C{a});(\C{l}_1, \C{b});(\C{l}_4, \C{b}); (\C{l}_2, \C{a})\}; \I{mem}(l, u) \equiv \{(\C{l}_0, \C{a});(\C{l}_1, \C{b});(\C{l}_4, \C{b}); (\C{l}_2, \C{a}); (\C{l}_2, \C{b})\}\}
\end{align*}
\endgroup Although they come from different sources, both samples
provide the values of variables (e.g., the value of $\nu_\S{is\_empty}$
in both $s^-$ and $s^+$ is $\bot$) and the values of predicate applications
(e.g., $\I{hd}(\nu_{\S{concat}}, \nu_{\S{top}})$ is true in $s^-$
sample and false in $s^+$).

Using $\llbracket \cdot \rrbracket$, we can \emph{extract} a
collection of feature vectors under a feature set $\mathcal{S}$ from a
sample $s$:
\begin{align*}
  \chi_\mathcal{S}(s) \equiv & \{\mathit{fv} \in 2^\mathcal{S}~|~ s \models \llbracket \mathit{fv} \rrbracket \}
\end{align*}
For example, the feature vectors extracted
from $s^-$ and $s^+$ for the feature set of $\S{top}$ are shown in the second and third rows of
\autoref{tab:neg} and \autoref{tab:pos}, respectively.

\begin{definition}[Classifier Consistency]
  \label{def:sampleconsistency}
  For a verification query $(\Sigma, \Phi)$, we say that a
  verification interface $\Delta$ is \emph{consistent} with a negative sample
  $s^-$ if \emph{at least one} of the library specifications in
  $\Delta$ classifies \emph{one or more} features extracted from that sample as
  negative:
  \begin{align*}
    \exists R_f(\vec{\alpha}, \nu) \in \Sigma,\;
    \exists \mathit{fv} \in \chi_{\mathit{FSet}(P, f(\vec{\alpha}) = \nu, \vec{u})}(s^-),
    \mathit{fv}\in \Delta(R_f)(\vec{a}, \nu)^-
  \end{align*}

  Similarly, $\Delta$ is consistent with a positive sample $s^+$ if
  \emph{all} specifications in $\Delta$ positively identify
  \emph{every} feature vector extracted from $s^+$:
  \begin{align*}
    \forall R_f(\vec{\alpha}, \nu) \in \Sigma,\;
    \forall \mathit{fv} \in \chi_{\mathit{FSet}(P, f(\vec{\alpha}) = \nu, \vec{u})}(s^+), \mathit{fv}\in \Delta(R_f)(\vec{a}, \nu)^+
 \end{align*}



\end{definition}

\textbf{Example.} The verification interface $\Delta$ from
\autoref{fig:consistent_solution} is consistent with ($s^-$), as the
specification of the $\S{top}$ function labels as negative the following
feature vector of $s^-$,
$\I{fv}^- \equiv \{\I{hd}(\S{s}_1, \S{u}) \mapsto \sf false;
\I{mem}(\C{s}_1, \S{u}) \mapsto \sf false; \nu_{\S{top}} = \S{u}
\mapsto \sf true\}$ to negative. Furthermore, $\Delta$ is also
consistent with all the feature vectors extracted from $s^+$.


\begin{theorem}
\label{theorem:safesolution}
For a given specification configuration
$((\Sigma, \Phi), P, F, \Gamma_{P}, \Gamma_{F})$ and verification
interface $\Delta$, $\Sigma[\Delta] \implies \Phi$ is valid iff
$\Delta$ is consistent with all negative samples $s^-$; $\Delta$ is a
consistent interface iff $\Delta$ is consistent with all positive
samples $s^+$ entailed by $\Gamma_{F}$ and $\Gamma_{P}$.
\footnote{\NV{All positive samples $s^+$ entailed by $\Gamma_{F}$ and
    $\Gamma_{P}$ means every positive sample consistent with the
    observations encoded by $\Gamma_{F}$ and
    $\Gamma_{P}$.}}\footnote{Proofs for all theorems are available
  \releasesource{}.}
\end{theorem}

\subsection{Learning Maximal Verification Interfaces}
\label{sec:sec:weakening}
While \autoref{theorem:safesolution} identifies the conditions under
which a verification interface is safe and consistent, it does not ensure
that it is maximal. We frame the search for a maximal
solution as a learning problem \NV{for a single function specification
assuming all other specifications are fixed}.

\begin{definition}[Weakest safe specification]
\label{def:weakest_safe}
\NV{For a given verification query $(\Sigma,\Phi)$ and safe and
  consistent verification interface $\Delta$, $\phi$ is the weakest safe specification of $f$ iff
\begin{enumerate}
    \item $\Delta[R_f \mapsto \phi]$ is safe
    \item For a given bound on the number of quantified variables $k$
      allowed in the specification of $f$, there is no other
      specification $\phi'$ with $k$ quantified variables that makes $\Delta[R_f \mapsto \phi']$ safe such that $\phi \implies \phi'$.
\end{enumerate}
}
\end{definition}

\begin{definition}[Sample with respect to library function]
\label{def:sample_wrt}
For a verification query $(\Sigma, \Phi)$, and safe verification
interface $\Delta$, a sample $s$ is positive (resp. negative) with
respect to library function $f$ when $s$ is positive
(resp. negative) and consistent with the specifications of all other library
functions in the domain of $\Delta$:
\begin{align*}
&s \models \Sigma[\Delta[R_f \mapsto \top]] \land \Phi\\
&s \models \Sigma[\Delta[R_f \mapsto \top]] \land \neg\Phi \quad\text{\NV{(resp.)}}
\end{align*}
\end{definition}

Ideally, the \NV{weakest safe specification} of $f$ would be able to positively classify every such positive sample.
Because this is not possible in general
due to the intrinsic granularity of the hypothesis space,  we must
limit ourselves to covering some subset of this space instead. The
weakening relation between classifiers can be viewed as the difference
between the sets of positive and negative feature vectors induced by
classifiers:

\begin{definition}[Weakening feature vector and samples]
\label{def:weakeningfv}
For a verification query $(\Sigma, \Phi)$ and safe verification
interface $\Delta$, a weakening feature vector $\mathit{fv}$
distinguishes between a weaker safe specification and $\Delta(R_f)$:
\begin{align*}
  \mathit{fv} \in \Delta(R_f)^- \text{ and } \Sigma[\Delta[R_f \mapsto \unitaryc{\I{fv}} \lor \Delta(R_f)]] \implies \Phi
\end{align*}
A sample $s$ is a weakening sample when $s$ is positive with respect
to $f$ and includes some weakening feature vector.  Intuitively, such
weakening samples can be used to safely generalize $f$.  If we cannot
find any weakening sample, then the specification must have converged
to a maximal one.
\end{definition}

\textbf{Example.} The sample
\begin{align*}
\{&\C{s}_1 \mapsto \C{l}_0; \C{s}_2 \mapsto \C{l}_1; \nu \mapsto \C{l}_2; \nu_{\S{top}} \mapsto \C{a}; \nu_{\S{tail}} \mapsto \C{l}_3; \nu_{\S{concat}} \mapsto \C{l}_4; \nu_{\S{is\_empty}} \mapsto \bot; \\
&\I{hd}(l, u)\equiv \{(\C{l}_0, \C{a});(\C{l}_1, \C{a});(\C{l}_2, \C{a});(\C{l}_4, \C{a})\}; \I{mem}(l, u) \equiv \{((\C{l}_0, \C{a});(\C{l}_1, \C{a});(\C{l}_2, \C{a});(\C{l}_4, \C{a})\}\}
\end{align*}
is a positive sample with respect to $\S{push}$ since it makes both
input stacks and the output stack contain $\C{a}$ and also have the
head element $\C{a}$, which entails $\textcolor{CBPurple}{\Phi_{\C{concat}}}$. It is also
consistent with all the specifications in the verification interface
from \autoref{fig:consistent_solution} outside of the one for $\S{push}$.
\NV{The feature vector $\I{fv}$ shown in \autoref{tab:weaken} is extracted}
from this sample, and it is not included in
$\Delta(R_{\S{push}})^+$.
Moreover,
$\Sigma[\Delta[R_{\S{push}} \mapsto \unitaryc{\I{fv}} \lor
\Delta(R_{\S{push}})]] \implies \Phi$, thus $\I{fv}$ is a weakening
feature vector for $\S{push}$, and the above sample is a weakening
sample for $\S{push}$.



\begin{theorem}
\label{theorem:pos_query}
For a given specification configuration
$((\Sigma, \Phi), P, F, \Gamma_{P}, \Gamma_{F})$ and safe verification
interface $\Delta$, \NV{$\Delta(R_f)$ is weakest safe specification of
  $f$ if and only if there are no weakening samples for $f$.}
\end{theorem}

\section{Algorithm}

\label{sec:algorithm}

\begin{algorithm}[t!]
  \Params{ Specification configuration $((\Sigma, \Phi), F, P, \Gamma_{F}, \Gamma_{P})$}
  \Output{ Maximal verification interface $\Delta$ or counterexample \textsc{Cex}}
  $\vec{u} \leftarrow \emptyset$\;
  \While{true}{
    \Match{$\I{SpecInfer}((\Sigma, \Phi), F, P, \Gamma_{F},
      \Gamma_{P}, \vec{u})$}{
      \Case{Fail $s^-$}{
        \Match{$\I{ExtractCex}(s^-)$}{
          \lCase{\textsc{Cex}}{
            \Return{\textsc{Cex}}
          }
          \lCase{None}{
            $\vec{u} \leftarrow \vec{u} \cup \I{FreshVariable}()$
          }
        }
      }
      \lCase{Fail None}{
        $\vec{u} \leftarrow \vec{u} \cup \I{FreshVariable}()$
      }
      \Case{$\Delta$}{
        \Repeat{$\Delta = \Delta_0$}{
          $\Delta_0 \leftarrow \Delta$\;
          \lFor{$f \in F$}{
            $\Delta(R_f) \leftarrow \I{Weaken}((\Sigma, \Phi), P, \Delta, f, \vec{u})$
          }
        }
        \Return{$\Delta$}\;
      }
    }
  }
  \caption{Multi-Abductive Inference Algorithm.}
  \label{algo:multi}
\end{algorithm}

Algorithm~\ref{algo:multi} presents the abductive inference procedure
depicted in \autoref{fig:workflow_simplified}. The algorithm maintains
a list of variables $\vec{u}$ that defines the hypothesis space it is currently exploring. The main loop of the algorithm
searches for a maximal verification interface in this hypothesis
space, starting from an initially empty set of variables (line 1). We
first try to use the $\I{SpecInfer}$ subalgorithm
(Algorithm~\ref{algo:consistent}) to infer a verification interface
that is safe and consistent with the specification configuration. If
inference fails with a sample, we search for corresponding inputs that
trigger a safety violation using random testing via the
$\I{ExtractCex}$ subroutine (line 8). If we are unable to find such
unsafe inputs, we extend $\vec{u}$ with a fresh variable (lines 7-8)
and restart the loop under this enhanced hypothesis space.  Otherwise,
$\I{SpecInfer}$ has found a safe and consistent verification
interface, $\Delta$, which we then refine to a maximal solution via a
weakening loop. This loop iteratively weakens the specification of
each library function in $\Delta$ using the $\I{Weaken}$ subalgorithm
(Algorithm~\ref{algo:single}), and returns $\Delta$ once it has
reached a fixed-point.


\subsection{Specification Inference}

\begin{algorithm}[hbt!]
\Params{Specification configuration $(\Sigma, \Phi), F, P, \Gamma_{F},
  \Gamma_{P}$, variables in hypothesis space $\vec{u}$}
\Output{Consistent safe verification interface or reports failure}
\lFor{$f \in F$}{
               $\pi_f,\omega_f \leftarrow \emptyset, \emptyset$
          }
$\Delta \leftarrow \{R_f \mapsto \forall \vec{u}, \top\}$\;
\While{$\I{true}$}{
  \While{$true$}{
    \Match{$\I{Sample}(\Delta(R_f), \Gamma_{F}, \Gamma_{P})$}{
      \lCase{None}{\Break}
      \Case{Some $s^+$}{
        \For{$f \in F$}{
          $\pi_f \leftarrow
          \I{FvecFromSample}(\I{FSet}(P,f(\vec{\alpha}_f) = \nu_f , \vec{u}), s^+)~~ \cup~~ \pi_f$\;
          $\omega_f \leftarrow \omega_f \setminus \pi_f$}
        \lFor{$f \in F$}{
            $\Delta(R_f) \leftarrow \I{Learner}(\pi_f, \omega_f)$
        }
      }
    }
  }
  \Match{$\I{Verify}((\Sigma[\Delta] \implies \Phi))$}{
            \Case{$\I{OK}$}{
              \Match{$\I{Verify}(\neg \Sigma[\Delta])$}{
                \lCase{$\I{OK}$ }{ \Return{Fail None} }
                \lCase{$\I{Sat}\ \_$}{ \Return{$\Delta$}}
              }
            }
            \Case{$\I{Sat}\ s^-$}{
              \lFor{$f \in F$}{
                $\omega'_f \leftarrow
                \I{FVecFromSample}(\I{FSet}(P, f(\vec{\alpha}_f) = \nu_f , \vec{u}),
                s^-) ~~\setminus~~ \pi_f$
              }
              \lIf{$\bigwedge\limits_{f\in F} \omega'_f = \emptyset$}{
                \Return{Fail $s^-$}
              } \lElse {
              \ZFor $f \in F$ \ZDo
                $\omega_f \leftarrow \omega_f \cup \omega'_f$
              }
              \lFor{$f \in F$}{
                $\Delta(R_f) \leftarrow \I{Learner}(\pi_f, \omega_f)$
          }
        }
      }
    }

\caption{Safe and Consistent Specification Inference ($\I{SpecInfer}$)}
\label{algo:consistent}
\end{algorithm}

Our data-driven multi-abduction specification inference algorithm
(\emph{SpecInfer}) given by Algorithm~\ref{algo:consistent} assumes
three key components: a property-based random sampler
$\mathit{Sample}$, a satisfiability checker $\mathit{Verify}$, and a
classifier learner $\mathit{Learner}$.  $\mathit{Sample}$ takes a
first-order formula as input and attempts to find a counterexample via
random sampling, returning data instances which violate this
specification based on these counterexamples. $\mathit{Verify}$
returns \textsf{OK} when given a valid formula, and produces an
assignment of variables and member predicates (i.e. a sample) that
invalidates its input formula otherwise. $\mathit{Learner}$ takes two
disjoint sets of feature vectors, $\pi$ and $\omega$, as inputs, and
returns a formula that classifies elements of $\pi$ and $\omega$ as
positive and negative, respectively. \name{} uses a decision tree
algorithm~\cite{ruggieri2002efficient} as the underlying learning
framework.

In more detail, for each library function $f,$ $\I{SpecInfer}$
maintains a pair of positive and negative sets of feature vectors
$\pi_f$ and $\omega_f$; these are initially empty (line $1$). The
algorithm holds the current candidate verification interface in the
variable $\Delta$, which initially stores trivial specifications for
each library function. $\I{SpecInfer}$ implements a two-step
refinement loop which first calls $\I{Sample}$ to test \NV{the current
  candidate verification interface $\Delta$ against $\Gamma_F$ and
  $\Gamma_P$ in order to identify any library function behaviors that
  are inconsistent with their current specifications (line 5). If so,
  $\I{Sample}$ returns a positive sample $s^+$ that we then use to augment
  the set of positive feature vectors (line 7). We subsequently remove any
  feature vectors appearing in $\pi_f$ from $\omega_f$ in order to
  ensure the two sets are disjoint before invoking $\I{Learner}$ to
  ascertain whether $\Delta$ is consistent with the newly observed behaviors
  (lines 9-11).}

\NV{$\I{SpecInfer}$ then queries $\I{Verify}$ to determine if the
  current verification interface is sufficient to ensure the desired
  safety property (line 12). We then verify that the specifications
  are not contradictory (line 14) before returning them, in order to
  ensure that they are not vacuously safe. Alternatively, if the verifier
  returns a sample $s^-$ witnessing an unsafe execution that $\Delta$
  was not strong enough to disallow, we extract sets of
  feature vectors for each library function $f$ from $s^-$, filtering
  out any that also occur in $\pi_f$ (line 18) to maintain the
  invariant that $\pi_f$ and $\omega_f$ are disjoint. If no new
  negative feature vectors are found, $\I{SpecInfer}$ cannot ensure
  that $s^-$ is spurious in the current hypothesis space, and it thus
  fails, returning $\textsf{Fail } s^-$ to the main
  algorithm. Otherwise, the algorithm updates $\omega_f$ with the new
  negative feature vectors and strengthens $\Delta$ to rule out $s^-$
  by invoking $\I{Learner}$ and the refinement
  loop restarts (line 21).}

\begin{theorem}[SpecInfer is Total and Sound]
\label{theorem:consistent-correct}
Algorithm~\ref{algo:consistent} always halts and returns a
verification interface that is safe and consistent.
\end{theorem}


\subsection{Weakening}

\begin{algorithm}[t!]
\Params{$(\Sigma, \Phi), P, \Delta, f, \vec{u}$}
\Output{Weakest safe specification of $f$}
$W_f \leftarrow \forall\vec{u}, \bot;\ \pi, \omega \leftarrow \emptyset, \emptyset $\;
\While{$\I{true}$}{
    \Match{$\I{Verify}(\textcolor{CBLightOrange}{\Sigma[\Delta[R_f \mapsto \forall
      \vec{u}, \top]] \land \Phi} \implies \Sigma[\Delta[R_f \mapsto
      \textcolor{CBPurple}{\Delta(R_f) \lor W_f} \lor \textcolor{CBOrange}{\unitaryc{\omega}}]])$}
    {
      \uCase{$\I{OK}$}{
        \leIf{$W_f = \forall\vec{u}, \bot$}{\Return{$\Delta(R_f)$}}{\Return{$W_f \lor \Delta(R_f)$}}
      }
      \Case{$\I{Sat}\ s$}{
        $W_f, \pi, \omega \leftarrow $ $\I{Update}$ $(s, W_f, \pi, \omega)$\;
        $W_f, \pi, \omega \leftarrow $ $\I{SafetyLoop}$ $(W_f, \pi, \omega)$\;
      }
    }
  }

\Procedure{$\I{Update}(s, W_f, \pi, \omega)$}{
    \For{$\I{fv} \in \I{FVecFromModel}(Fset(P, f(\alpha_f) = \nu_f, \vec{u}),\; s)$}{
        \Match{$\I{Verify}(
        (\Sigma[\Delta[R_f \mapsto W_f \lor \Delta(R_f) \lor \unitaryc{\I{fv}}]] \implies \Phi))$}{
            \lCase{$\I{OK}$}{
                $\pi \leftarrow \pi \cup \{\I{fv}\}$
            }
            \lCase{$\I{Sat}\ \_$}{
                $\omega \leftarrow \omega \cup \{\I{fv}\}$
            }
          }
          $W_f \leftarrow \I{Learner}(\pi, \omega)$
    }
    \Return{$W_f, \pi, \omega$}
}

\Procedure{$\I{SafetyLoop}(W_f, \pi, \omega)$}{
    \Match{$\I{Verify}((\Sigma[\Delta[R_f \mapsto W_f \lor \Delta(R_f)]] \implies \Phi))$}{
        \lCase{$\I{OK}$}{
            \Return{$W_f, \pi, \omega$}
        }
        \Case{$\I{Sat}\ s$}{
            $W_f, \pi, \omega \leftarrow \I{Update}(s, W_f, \pi, \omega)$\;
            \Return{$\I{SafetyLoop}(W_f, \pi, \omega)$}
        }
    }
}
\caption{Weakening ($\I{Weaken}$)}
\label{algo:single}
\end{algorithm}

Algorithm~\ref{algo:single} shows the $\I{Weaken}$ procedure that is
used to generalize the specification of a library function $f$ from the
initial solution returned by $\I{SpecInfer}$.  This procedure takes a
verification query, method predicate set, verification interface,
library function, and a list of quantified variables as input, and
infers a weakest safe specification for $f$. Notably, the arguments of
$\I{Weaken}$ do not include $\Gamma_P$ or $\Gamma_P$, reflecting that
our approach to weakening is purely logical.

$\I{Weaken}$ maintains a formula $W_f$ which is used to weaken the
specification of $f$; the start of the function initializes this
formula to false (line 1). Like $\I{SpecInfer}$, $\I{Weaken}$
maintains two sets of feature vectors: $\pi$, which holds weakening
feature vectors, and $\omega$, which holds feature vectors
representing library behaviors that cause client safety
violations. Both sets are initially empty (line 1).  The body of
$\I{Weaken}$ uses a counterexample-guided refinement loop to
iteratively construct $W_f$. The loop first queries the verifier to ensure that
every \textcolor{CBLightOrange}{safe execution consistent with the candidate
specifications besides $\Delta(R_f)$} is either \textcolor{CBPurple}{a)
currently allowed by $W_f \lor \Delta(R_f)$} or \textcolor{CBOrange}{b) involves
a behavior of $f$ that can trigger a safety violation in other contexts (line
3).} The clauses of the formula representing each of these pieces are highlighted
with the corresponding color.
When this query is valid, there is no weakening sample for $f$; thus,
by \autoref{theorem:pos_query}, the current specification for $f$ is
the weakest safe one, and $\I{Weaken}$ returns $W_f \lor \Delta(R_f)$
(or $\Delta(R_f)$ when $W_f \equiv \bot$) (line 5).
Alternatively, \emph{Verify} will provide a sample $s$ witnessing a
safe execution; this sample is passed to the $\I{Update}$ subroutine
to further weaken $W_f$. $\I{Update}$ first uses $\I{Verify}$ to identify
feature vectors extracted from $s$ that trigger safety violations in
other contexts. After inserting safe and unsafe feature vectors into
$\pi$ and $\omega$, respectively (lines 12-13), $\I{Update}$ uses
$\I{Learner}$ to refine $W_f$.
The weakened specification $W_f \lor \Delta(R_f)$ may violate client
program safety, however. We use a learning based
procedure for its intrinsic generalization ability: when we add a new
feature vector to $\pi$, the learned classifier will not only classify
\emph{that} new feature vector to positive, but also potentially
others. This ability results in faster convergence, since we do not need
to explicitly enumerate each weakening feature vector. However, to
avoid the learned classifier from being weakened to the point
that it mistakenly classifies unsafe feature vectors as safe,
$\I{Weaken}$ then invokes $\I{SafetyLoop}$ to filter any unsafe
weakenings (line 8).
\NV{$\I{SafetyLoop}$ implements a similar
  strategy as Algorithm~\ref{algo:consistent} to progressively refine
  $W_f$ until it guarantees safety. After this step, the loop restarts
  in order to identify additional weakening possibilities}.


\begin{theorem}[Weaken is Sound]
\label{theorem:weakening-correct}
For given $(\Sigma, \Phi), P, \Delta, f, \vec{u}$, if $\Delta$ is
safe, Algorithm~\ref{algo:single} will halt with the weakest safe
specification for $f$.
\end{theorem}


\begin{theorem}[Relative Completeness]
\label{theorem:multi-correct}
Algorithm~\ref{algo:multi} either returns either a maximal verification interface or a concrete
counterexample.
\end{theorem}

\section{Implementation and Evaluation}
\label{sec:evaluation}

We have implemented a verification pipeline based on the above
approach, called \name{}\cite{zhe_zhou_2021_5130646}, that targets OCaml programs which rely on
libraries to manipulate algebraic data types. \name{} consists of
$7267$ lines of OCaml and uses Z3~\cite{de2008z3} as its backend
solver. \name{}'s frontend generates verification queries from client
programs via weakest liberal precondition predicate
transformers. \name{} does not automatically infer inductive
invariants for recursive client functions, and it expects client
programs to provide such invariants, in addition to any pre- and
post-conditions.

\begin{table}[]
\renewcommand{\arraystretch}{0.8}
\begin{tabular}{ccc|ccc|r@{$~/~$}lcc}
\toprule
                        & $(|F|, |R|)$ & $|P|$ & $|\vec{u}|$ & $|cex|$ & time$_{c}$(s) & \#Gather&$|\phi^+|$ & time$_{w}$(s) & time$_{d}$(ms) \\
\midrule
\multirow{4}{*}{queue}  & $(6, 10)$   & $2$ & $2$         & $26$    & $1.85$   & $366$&$1828$ & $47.9$ & $123.5$ \\
                        & $(4, 5)$   & $2$ & $1$         & $10$    & $0.28$    & $235$&$1536$ & $9.2$ & $18.2$ \\
                        & $(4, 5)$   & $3$ & $2$         & $23$    & $0.87$    & $2076$&$30054$ & $318.9$ & $42.7$\\
                        & $(6, 10)$   & $1$ & $1$         & $8$     &
                                                                      $0.22$   & $53$&$114$ & $1.9$ & Max \\
\midrule
\multirow{3}{*}{stack}  & $(4, 5)$   & $2$ & $1$         & $11$    & $0.61$    & $29$&$13$ & $0.5$ & $18.5$ \\
                        & $(4, 5)$   & $3$ & $2$         & $39$    & $2.94$    & $3220$&$22716$ & Limit & $169.3$ \\
                        & $(4, 5)$   & $2$ & $1$         & $4$     & \textcolor{red}{$0.08$}  & \multicolumn{3}{c}{}\\

\midrule
\multirow{5}{*}{heap}   & $(3, 8)$   & $2$ & $2$         & $71$    & $11.21$   & $1790$&$26588$ & Limit & $895.8$\\
                        & $(2, 21)$   & $1$ & $1$         & $10$    & $0.37$    & $172$&$432$ & $21.5$ & $21.3$\\
                        & $(2, 21)$   & $3$ & $2$         & $141$   & $107.18$  & $702$&\textcolor{blue}{$177388$} & Limit & $328.3$\\
                        & $(2, 21)$   & $3$ & $2$         & $192$   & $152.99$  & $512$&\textcolor{blue}{$190215$} & Limit & $149.7$\\
                        & $(3, 8)$   & $1$ & $1$         & $10$    &
                                                                     $0.37$    & $464$&$3308$ & $46.0$ & Max\\
\midrule
\multirow{4}{*}{stream} & $(4, 10)$   & $1$ & $1$         & $4$     & $0.15$    & $50$&$110$ & $1.3$ & $21.1$\\
                        & $(4, 10)$   & $2$ & $2$         & $23$    & $1.57$   & $640$&$1376$ & $1038.0$ & $70.6$ \\
                        & $(4, 10)$   & $3$ & $2$         & $18$    & $1.69$  & $1755$&$31276$ & $741.8$ & $24.3$\\
                        & $(4, 10)$   & $3$ & $2$         & $12$    & \textcolor{red}{$0.37$}  &\multicolumn{3}{c}{}\\
\midrule
\multirow{9}{*}{set}    & $(1, 21)$   & $4$ & $2$         & $91$    & $77.75$  & $1369$&\textcolor{blue}{$59797035$} & Limit & $1268.7$\\
                        & $(2, 8)$   & $1$ & $1$         & $11$    & $0.30$    & $181$&$448$ & $11.2$ & $21.1$\\
                        & $(2, 8)$   & $2$ & $1$         & $21$    & $0.81$    & $3327$&$27609$ & $1741.9$ & $23.2$\\
                        & $(2, 8)$   & $2$ & $2$         & $63$    & $8.28$    & $2496$&$28495$ & Limit & $314.1$ \\
                        & $(2, 6)$   & $3$ & $1$         & $15$    & $0.39$   & $3644$&$22526$ & $1203.3$ & $19.5$\\
                        & $(2, 6)$   & $3$ & $1$         & $25$    & $0.69$   & $3091$&$20090$ & $968.7$ & $19.2$ \\
                        & $(1, 21)$   & $1$ & $1$         & $10$    &
                                                                      $0.27$   & $228$&$772$ & $34.0$ & Max \\
                        & $(2, 8)$   & $2$ & $1$         & $8$     & \textcolor{red}{$0.14$}  & \multicolumn{3}{c}{}\\
                        & $(2, 6)$   & $3$ & $3$         & $29$    & \textcolor{red}{$4.69$}   & \multicolumn{3}{c}{}\\
\midrule
\multirow{2}{*}{trie}   & $(4, 18)$   & $2$ & $1$         & $12$    & $0.47$  & $167$&$404$ & $13.6$ & $28.0$\\
                        & $(4, 18)$   & $2$ & $1$         & $15$    & \textcolor{red}{$0.31$} & \multicolumn{3}{c}{}\\
\bottomrule
\end{tabular}
\caption{\small Experimental results. The columns in the table can be
  divided into three groups. The first group presents the number of
  distinct library functions ($|F|$), the number of library function
  applications ($|R|$), and the size of the method predicate set
  ($|P|$) for each benchmark. The second column describes the number
  of quantified variables ($|\vec{u}|$), the number of
  counter-examples generated ($|cex|$), and the time in seconds
  (time$_{c}$) needed for \name{} to find a consistent and safe
  verification interface. Times indicate how long it took for
  counter-example generation, sampling, feature vector extraction,
  and labelling and DT learning. They do not include the time taken to
  generate the initial verification query. \textcolor{red}{Red}
  entries in the time column indicate how long it took to identify
  safety violations in the five unsafe benchmarks. The last group
  lists the number of gathered (\#Gather) and total positive
  ($|\phi^+|$) feature vectors in the space of weakenings, the time
  needed by the weakening phase (time$_{w}$), and the time needed for
  the SMT solver to find a sample allowed by a weakened solution but
  not the initial one (time$_{d}$). \textcolor{blue}{Blue} entries in
  the $|\phi^+|$ column indicate a lower bound. ``Limit'' entries in
  the time$_{w}$ column indicate that the one hour time bound was
  reached. ``Max'' entries in the time$_{d}$ column indicate that the
  initial solution was already maximally weak.}
\label{tab:evaluation}
\end{table}

Our experimental evaluation of \name{} addresses five key
questions:
\begin{itemize}
\item[\textbf{Q1}:] Is \name{} able to find specifications sufficient
  to verify a range of properties and client programs in a reasonable
  amount of time?
\item[\textbf{Q2}:] Can \name{} identify unsafe client programs?
\item[\textbf{Q3}:] Can \name{} efficiently find maximal solutions?
\item[\textbf{Q4}:] Is \name{} able to find useful intermediate
  generalizations of initial specifications?
\item[\textbf{Q5}:] Does weakening improve the quality of the
    inferred specifications?
\end{itemize}

\noindent All reported data was collected on a Linux server
with an Intel(R) Core(TM) i7-8700 CPU $@$ 3.20GHz and 64GB of RAM.

To answer these questions, we have evaluated \name{} on a
corpus\footnote{All benchmarks,
  post-conditions, and inferred library specifications from our
  evaluation are provided \releasesource{}.} of client programs drawn
from~\citet{okasaki1999purely}, the OCaml standard
library~\cite{leroy2014ocaml}, Verified Functional
Algorithms~\cite{appel2018software} and Software
Foundations~\cite{pierce2010software}. Our benchmarks cover a range of
abstract data types manipulating a diverse set of algebraic data
types, including queues (bankers queues and batched queues),
list-based stacks, heaps (leftist heaps and splay heaps), streams, sets
(backed by trees, red black trees, and lists), and tries. The underlying
representation of each algebraic data type provides a set of method
predicates related to ordering, membership, and uniqueness (no
duplicate elements) that can be queried as part of a test. \NV{These
  sorts of shape properties are relatively under-constrained and are thus amenable to property-based random sampling.}  We used
\name{}\ to verify several different properties for each data type,
including membership, ordering, distinct elements, and sorting.
\begin{example}
  One of our benchmarks uses the following specification for an
  $\S{insert}$ program that inserts an element $\C{x}$ into an
  unbalanced set $\C{s}$ using a binary tree for its underlying
  representation:
\begin{align*}
    (\forall u, &(\I{root}(\C{s}, u) \implies (u < \C{x} \implies \I{root}(\C{\nu_\S{insert}}, u)) \land (u \geq \C{x} \implies \I{root}(\C{\nu_\S{insert}}, \C{x}))) \\
  \land~~ &(\I{mem}(\C{\nu_\S{insert}}, u) \iff \I{mem}(\C{s}, u) \lor \C{x} = u)
\end{align*}
Given this specification, \name{} infers the following specification
for the $\S{maket(x, l, r) = \nu}$ function used by $\S{add}$ that constructs a new
tree from $\C{x}$ and the left and right subtrees $\C{l}$ and $\C{r}$:
\begin{align*}
  \forall u, &(\I{mem}(\C{\nu}, u) \iff \I{mem}(\C{l}, u) \lor \I{mem}(\C{r}, u) \lor (\C{x} = u)) \land (\I{root}(\C{\nu}, u) \iff \C{x} = u) \\
  \land &(\I{root}(\C{l}, u) \implies \I{mem}(\C{l}, u)) \land (\I{root}(\C{r}, u) \implies \I{mem}(\C{r}, u))
\end{align*}
\noindent The first line of this specification captures the key
semantic properties of the tree, while the second line encodes a key
relationship between the $\I{mem}$ and $\I{hd}$ method predicates
needed by the solver to verify the specification.
\end{example}



The detailed results of our experiments are shown in
\autoref{tab:evaluation}. The first group of columns in
\autoref{tab:evaluation} describes the salient features of our
benchmarks. Each client specification uses between $1$ and $4$ member
predicates, and the client programs make between $5$ and $21$ calls to
library functions.  In order to evaluate \name{}'s ability to identify
faulty specifications, our experiments also included five unsafe
client programs. \name{} can infer specifications for all the
benchmarks with valid assertions, and returns concrete
counter-examples for clients with unsafe post-conditions.

The second group of columns in \autoref{tab:evaluation} presents our
evaluation of \name{}'s ability to discover an initial safe and
consistent verification interface (\textbf{Q1} and \textbf{Q2}). These
columns show that \name{} is relatively efficient at finding safe
specifications, with none of the benchmarks taking longer than three
minutes to learn an initial solution (\textbf{Q1}). As expected, more
complicated benchmarks (i.e. those with more function calls or member
predicates) required more SMT-provided counterexamples and took longer
to complete, as did benchmarks requiring a larger hypothesis space
(i.e. specifications with more quantified variables). Notably, \name{}
was able to quickly generate concrete counter-examples witnessing
safety violations (\textbf{Q2}) in the five benchmarks with
invalid postconditions (these benchmarks have \textcolor{red}{red}
entries in the time$_{c}$ column).

The final group of columns in \autoref{tab:evaluation} addresses the
questions dealing with \name{}'s weakening phase (\textbf{Q3 -
  Q5}). Unsurprisingly, weakening requires more time than the initial
inference phase; while the latter needs to identify a \emph{single}
solution, the former needs to account for \emph{all} possible
solutions in order to select the best one. When evaluating this phase,
we choose to use a one hour time bound for each experiment. If this
time bound was reached, we had \name{} return the current weakened
solution. Three-fourths (16/22) of our safe benchmarks were able to
find maximal solutions from the initial solution within this limit
(\textbf{Q3}).  In general, the time taken to weaken a solution was
correlated with the complexity of the benchmark.


To further investigate how effective \name{} was at exploring the
space of candidate weakenings (\textbf{Q4}), we calculated the ratio
of the total number of feature vectors gathered during weakening
(\#Gather) against the total number of positive feature vectors
admitted by the final maximal specification ($|\phi^+|$). The latter
number represents the set of vectors that a na\"{i}ve exhaustive
enumeration would need to consider in order to find our solution. In
our experiments, \name{} only needed to consider at most $40\%$ of the
full search space.\footnote{In the first stack benchmark, \name{}
  gathers more feature vectors than the total number because the
  algorithm may generate false feature vectors, as discussed in
  \autoref{sec:algorithm}.} To get $|\phi^+|$ for the six benchmarks
on which \name{} returned a partially weakened solution, we increased
the time bound to 24 hours and ran \name{} until it converged on the
maximal solution. \name{} was unable to converge under this longer
bound for three of these benchmarks. For these three, we report the
total number of feature vectors in the partial solution, which is
indicated using a \textcolor{blue}{blue} entry in the $|\phi^+|$
column.

In an attempt to quantify how well \name{} was able to generalize an
initial solution, we asked the SMT solver to identify samples
permitted under the weakened solution but not the original
(\textbf{Q5}). The more general a weakened solution, the larger this
space should be, so the solver should be able to more readily identify
one of its elements. The results of this experiment are presented
under the time$_{d}$ column. The initial solution was already maximal
for three of our benchmarks, which is indicated via a ``Max'' entry in
the time$_d$ column. In general, the solver was able to quickly find
such samples for the remaining nineteen benchmarks. This search took
longer for the benchmarks that hit the time bound, suggesting those
solutions are either closer to the initial specifications or are
otherwise more complicated for the SMT solver to handle.

\begin{figure}[ht]
  \includegraphics[width=400pt]{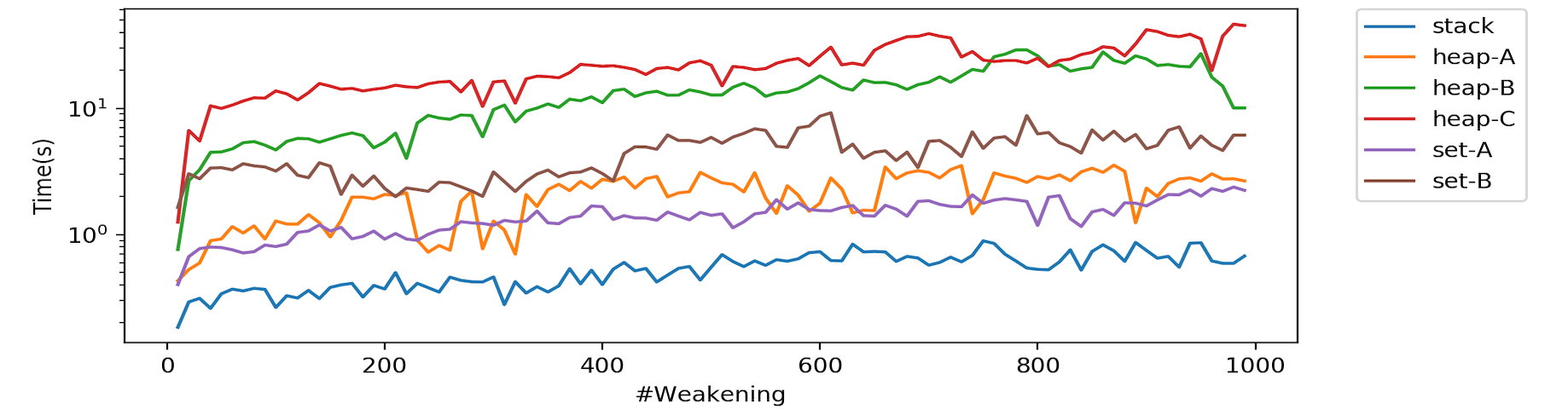}
\centering
\caption{\small Run times of the first $1000$ weakening iterations in benchmarks.}
\label{fig:time_of_each_weakening}
\end{figure}
When running our experiments without a time bound, we observed that
individual iterations of the weakening loop took longer to complete
over time. As an example, in the first hour of the second stack
benchmark, \name{} performed $46$ weakening steps per minute on
average, but only averaged $33$ after that.  Based on this
observation, we conjectured that it became harder for \name{} to find
further generalizations as the weakening phase progresses. To test
this hypothesis, we measured the time required by each of the first
$1000$ iterations of the weakening loop for the six benchmarks that
hit the time bound. The results of this experiment are presented in
\autoref{fig:time_of_each_weakening}; the y-axis uses a logarithmic
scale in order to account for times that range from $0.1$s to $50$s.
The trajectory agrees with our hypothesis, and suggests diminishing
returns for running the weakening procedure for long periods of time.


Finally, in order to show that \name{} produces useful specifications
(\textbf{Q1} and \textbf{Q5}), we used the Coq proof assistant to
verify that implementations of the library functions from our
benchmarks satisfied their inferred specifications.\footnote{The
  corresponding Coq developments are provided \releasesource{}.} In total, \name{} inferred $68$ specifications across
all our experiments; we were able to verify $64$ of these in Coq. Of
these, three had initial specifications that were too strong; it was
only after the weakening phase that the specifications could be used
for verification. The four specifications that we were not able to
verify came from the benchmarks whose weakening phase timed out,
i.e., these specifications were not maximal.  Taken together, these
points suggest that the weakening phase results in more general (and
thus more useful) specifications, and that \name{} is highly effective in
finding meaningful specifications.

\section{Related Work}
\label{sec:related}

\paragraph{Data-Driven Approaches} There have been several recent
data-driven approaches that use learning to infer specifications of
library functions. \citet{zhu2016automatically} automatically infer
specifications that use a fixed set of features (analogous to our
method predicates) to identify relationships between the input and
outputs of a function.  \citet{padhi2016data} use program synthesis to
automatically learn features on demand when inferring preconditions
for data-structure manipulating library functions. This approach is
further extended by~\citet{RepInvGen} to synthesize~\cite{OZ15}
representation invariants that are sufficient to verify specifications
of the abstract datatype operations.  Rather than inferring
specifications of libraries in isolation, we consider the
complementary problem of discovering those specifications in service
of verifying library clients. \NV{Broadly speaking, this change in
  perspective differentiates our approach from these prior works
  insofar as we cannot appeal to any source of ground truth (e.g, SMT
  verification as in \citet{padhi2016data} or bounded model checking
  as in \citet{RepInvGen}).  This leads to the need for a weakening
  procedure to facilitate generalization to ultimately aid client-side
  verification.  However, techniques like
  \citet{RepInvGen} could be used alongside ours, for example, by
  leveraging abduced specifications to infer and verify representation
  invariants.}

\paragraph{Automated Verification} Encoding verification conditions in
a logic for which efficient solvers exist (e.g. SMT) is ubiquitous in
the automatic program verification community. In this setting, the
standard approach to reasoning about clients of user-defined functions
is to rely on some manually written axiomatization of those functions,
paying particular care in order to ensure that the underlying solver
will terminate~\cite{itzhaky2014modular, itzhaky2013effectively}. In
the case that a specification is incomplete, e.g., when defining new
functions, manual intervention is required to extend the
axiomatization. Informally, our approach can be thought of as filling in the
missing parts of specifications by using the latent semantics of
method predicates. More recently, \citet{RefinementReflection}
introduced \emph{refinement reflection} in order to enable SMT-based
reasoning about arbitrary user-defined functions. There, the semantics
of a function are embedded directly into the logic as a set of
equations, and users can \emph{manually} construct equational proofs
about their behavior using a library of proof combinators. The authors
introduce a proof search algorithm to help automate the construction
of these proofs. While this algorithm is complete when a proof exists
for a bounded unfolding of function definitions, users are still
required to provide instantiations of lemmas and induction
hypotheses to completely automate program verification.  Our approach
uses data-driven methods and counterexample-guided search to generate
specifications without the need to reflect the implementation, which
in our setting is unavailable, into the solver's underlying logic.

\paragraph{Abductive Inference} As noted in \autoref{sec:motivation},
our logical formulation of specification inference is an instance of
an abductive inference problem. This observation has been previously
exploited to develop inference algorithms for loop
invariants~\cite{DilligAbductiveInv} and specifications of functions
in a client program~\cite{AlbarghouthiMSS}. For a given program, both
algorithms rely on an abduction procedure to iteratively strengthen
loop invariants (resp. function specifications) until they are
strong enough to prove a user-provided post-condition. While
completely automated, these approaches critically rely on an abduction
procedure for the underlying specification logic, in particular the
first-order theory of linear integer arithmetic in their experiments.
To the best of our knowledge, no such abduction procedure exists for
the theory of equalities with uninterpreted function symbols that is
commonly used to specify recursive functions over algebraic
datatypes. Our approach provides an alternative solution that combines
data-driven methods with SMT-based counterexample-guided refinement to
discover library method specifications to aid
client-side verification.

\NV{While \citet{bastani2015specification} consider a
  similar abductive inference problem -- discovering minimal sufficient
  assumptions  in order to analyze client programs when some
  portion of the code is missing -- there are several major differences between our two
  approaches: 1) their specification domain is limited to CFL
  reachability (alias and taint specifications), while our hypothesis
  space involves shape properties (membership, ordering, etc.)  over
  algebraic datatypes; 2) we guarantee specification consistency with
  respect to our observations on the blackbox implementation of the
  libraries; and (3) our approach incorporates an explicit weakening
  procedure to generalize candidate solutions.}

\NV{\paragraph{Specification Inference} Similar to our problem setup,
  \citet{nguyen2014mining} and \citet{su2018using} infer
  specifications of library APIs from client information. However,
  both works try to infer specifications of correct library usage
   under the assumption that the client is itself correct. In
  contrast, we leverage abductive methods to provide guarantees
  that any library implementation must satisfy to ensure the client is
  safe.  Our approach does not assume clients are always safe -- e.g,
  \name{} was able to identify safety violations in $5$ of our
  benchmarks.  Another distinguishing feature of our work is our focus
  on generating maximally weak specifications to overcome
  overfitted specifications. The work of \citet{pandita2012inferring}
  learns specifications from comments in natural language but does not
  provide safety and consistency guarantees.}

\NV{\citet{qin2010verifying} also infers specifications for unknown
  procedures in a rich domain involving shape properties of
  datatypes. However, their technique does not query library
  implementations and does not generalize inferred specifications,
  which can potentially lead to specifications that are overfitted to
  the client safety property.  In contrast to our work, their method
  also requires users to understand the underlying representation of
  the datatype used by the library method, and present explicit
  interpretations of predicates used in specifications sufficient to
  verify the client.  Finally, we apply an additional algorithmic
  weakening procedure to refine inferred specifications.}

\NV{\paragraph{CHC solving} While it is possible to frame our problem as
  an instance of data-driven CHC solving \`{a} la \citet{ZMJ18}, this
  would not address the important challenge of generating
  maximally weak specifications, a concern that is not considered by
  Horn-clause semantics~\cite{AlbarghouthiMSS}.}

\section{Conclusions}
\label{sec:conc}
This paper presents a novel data-driven approach to \NV{infer
  specifications using a minimal set of assumptions that are
  nonetheless consistent with provided blackbox library
  implementations and sufficient to verify client assertions}.  We
demonstrate that our technique, manifested in a tool called \name{},
is highly effective in identifying sophisticated \NV{specifications}
that enable verification of challenging functional data structure
programs.


\section*{Acknowledgements}

We thank Pedro Abreu and the anonymous reviewers for their detailed comments and suggestions.
This material is based upon work supported by the NSF under Grants
CCF-SHF 1717741, CCF-FMiTF 2019263, and CCF-1755880.

\bibliographystyle{ACM-Reference-Format}
\bibliography{bibliography}

\ifnum\pdfstrcmp{\rsource}{arxiv}=0
  \newpage
\appendix
\section{Benchmarks and Results}

An artifact containing this tool, our benchmark suite, results and corresponding Coq proofs is publicly available on \href{https://doi.org/10.5281/zenodo.5130646}{Zenodo:10.5281/zenodo.5130646}~\cite{zhe_zhou_2021_5130646}.

\section{Proofs of Theorems}

Here we clarify the definition of "quantified variables" from the previous sections. The quantified variables are divided into universally quantified variables and existentially quantified variables. Correspondingly, the "satisfying instantiation of quantified variables" also has two components: when a variable is universally quantified, a satisfying instantiation has to cover all possible values; when one variable is existentially quantified, the satisfying instantiation should provide a specific value in the domain.

\subsection{Lemma Specification Classification}

\begin{lemma}[Specification Classification]
For given classifier $\phi$,
\begin{align*}
    \bigvee_{\I{fv} \in \phi^+} \unitaryc{\I{fv}} \iff \phi 
\end{align*}
\label{lemma:specclassificatioin}
\end{lemma}
\textbf{Proof. }
According to Def \ref{def:classification}, $\I{fv} \in \phi^+$ means $\unitaryc{\I{fv}} \implies \phi$, thus $\bigvee_{\I{fv} \in \phi^+} \unitaryc{\I{fv}} \implies \phi$. On the other hand, if $\phi \centernot\implies \bigvee_{\I{fv} \in \phi^+} \unitaryc{\I{fv}}$ then $\bigvee_{\I{fv} \in \phi^+} \unitaryc{\I{fv}} \implies \neg \phi$. Thus there exists a feature vector $\I{fv}$ in $\phi^+$, $\I{fv} \centernot\implies \phi$ which is conflict with Def \ref{def:classification}. By contradiction, $\phi \implies \bigvee_{\I{fv} \in \phi^+} \unitaryc{\I{fv}}$.
\textbf{Qed. }

\subsection{Theorem \ref{theorem:safesolution}}
For a given specification configuration
$((\Sigma, \Phi), P, F, \Gamma_{P}, \Gamma_{F})$ and verification
interface $\Delta$, $\Sigma[\Delta] \implies \Phi$ is valid iff $\Delta$ is consistent with all negative samples $s^-$; $\Delta$ is a consistent interface iff $\Delta$ is consistent with all positive samples $s^+$ entailed by $\Gamma_{F}$ and $\Gamma_{P}$.

\textbf{Proof. } We first prove if $\Sigma[\Delta] \implies \Phi$, then $\Delta$ is consistent with all negative samples. Consider $\Sigma[\Delta]$ which is the conjunction of application of library specifications, according to Lemma \ref{lemma:specclassificatioin}:
\begin{align*}
   \Sigma[\Delta] \equiv \bigwedge_{R_f(\vec{\alpha}, \nu) \in \Sigma}, \bigvee_{\I{fv} \in \Delta(R_f)(\vec{\alpha}, \nu)^+} \unitaryc{\I{fv}}
\end{align*}
If there exists a negative sample $s^-$ which is not consistent with $\Delta$, by Def \ref{def:sampleconsistency}:
  \begin{align*}
    \forall R_f(\vec{\alpha}, \nu) \in \Sigma,\;
    \forall \mathit{fv} \in \chi_{\mathit{FSet}(P, f(\vec{\alpha}) = \nu, \vec{u})}(s^-),
    \mathit{fv}\not\in \Delta(R_f)(\vec{a}, \nu)^-
  \end{align*}
then
  \begin{align*}
    \forall R_f(\vec{\alpha}, \nu) \in \Sigma,\;
    \forall \mathit{fv} \in \chi_{\mathit{FSet}(P, f(\vec{\alpha}) = \nu, \vec{u})}(s^-),
    \unitaryc{\mathit{fv}} \implies \Delta(R_f)(\vec{a}, \nu)
  \end{align*}
which implies
\begin{align*}
    \forall R_f(\vec{\alpha}, \nu) \in \Sigma,\;
    s \models \Delta(R_f)(\vec{a}, \nu)
\end{align*}
that is $s \models \Sigma[\Delta]$. As $\Delta$ is safe, then $\Sigma[\Delta] \implies \Phi$, then $s \models \Phi$ which is conflict with $s$ is a negative sample. By contradiction, $\Delta$ is consistent with all negative samples.

Then we prove consistent verification interface $\Delta$ is consistent with all positive samples $s^+$ entailed by $\Gamma_{F}$ and $\Gamma_{P}$. If there exists a positive sample $s^+$ entailed by $\Gamma_{F}$ and $\Gamma_{P}$ which is not consistent with $\Delta$, by Def \ref{def:sampleconsistency}:
  \begin{align*}
    \exists R_f(\vec{\alpha}, \nu) \in \Sigma,\;
    \exists \mathit{fv} \in \chi_{\mathit{FSet}(P, f(\vec{\alpha}) = \nu, \vec{u})}(s^+),
    \mathit{fv}\not\in \Delta(R_f)(\vec{a}, \nu)^+
  \end{align*}
As $s^+$ is entailed by $\Gamma_{F}$ and $\Gamma_{P}$, then there exists a concrete value assignment which provides the input and output of each function call, specifically, the input and output of $f$ is $\vec{\alpha}_{s^+}$ and $\nu_{s^+}$:
\begin{align*}
    s^+ \models \mathit{FSet}(P, f(\vec{\alpha}_{s^+}) = \nu_{s^+}, \vec{u})[\Gamma_P]
\end{align*}
According to Def \ref{def:mapconsistency}:
\begin{align*}
    \mathit{FSet}(P, f(\vec{\alpha}_{s^+}) = \nu_{s^+}, \vec{u})[\Gamma_P] \implies \Delta(R_f)(\vec{\alpha}_{s^+},\nu_{s^+})[\Gamma_P]
\end{align*}
then
\begin{align*}
    s^+ \models \Delta(R_f)(\vec{\alpha}_{s^+},\nu_{s^+})[\Gamma_P]
\end{align*}
On the other hand all feature vectors in $s^+$ is
\begin{align*}
    \chi_{\mathit{FSet}(P, f(\vec{\alpha}) = \nu, \vec{u})}(s^+) \equiv 
     \{\mathit{fv} ~|~ s^+ \models \unitaryc{\I{fv}}\}
\end{align*}
which means
\begin{align*}
 \forall \mathit{fv}, s^+ \models \unitaryc{\I{fv}} \implies \I{fv} \in \chi_{\mathit{FSet}(P, f(\vec{\alpha}) = \nu, \vec{u})}(s^+)
\end{align*}
Then all feature vectors consistent with $\Delta(R_f)$ is in $\chi_{\mathit{FSet}(P, f(\vec{\alpha}) = \nu, \vec{u})}(s^+)$ which is conflict with $\Delta$ is not consistent with $s^+$. By contradiction, $\Delta$ is consistent with all positive samples.

Now we prove, if a verification interface $\Delta$ is consistent with all negative samples then $\Sigma[\Delta] \implies \Phi$. As $\Delta$ is consistent with all negative $s^-$, then by Def \ref{def:sampleconsistency}:
  \begin{align*}
    \exists R_f(\vec{\alpha}, \nu) \in \Sigma,\;
    \exists \mathit{fv} \in \chi_{\mathit{FSet}(P, f(\vec{\alpha}) = \nu, \vec{u})}(s^-),
    \mathit{fv} \in \Delta(R_f)(\vec{a}, \nu)^-
\end{align*}
Then $s^- \not\models \Sigma[\Delta]$, then $\Sigma[\Delta] \implies \Phi$.

Finally we prove, if a verification interface $\Delta$ is consistent with all positive samples $s^+$ entailed by $\Gamma_{F}$ and $\Gamma_{P}$, it is consistent. If $\Delta$ is not a consistent interface, by Def \ref{def:mapconsistency}, there exists a library function $f$ and one sample $s$ entailed by $\Gamma_{F}$ and $\Gamma_{P}$ which assigns the input and output of $f$ to $\vec{\alpha}_s$ and $\nu_s$,
\begin{align*}
    s \not\models \mathit{FSet}(P, f(\vec{\alpha}_{s}) = \nu_{s}, \vec{u})[\Gamma_P]
\end{align*}
As the verification queries is correct, then all samples entailed entailed by $\Gamma_{F}$ and $\Gamma_{P}$ are positive, then $s$ is the positive which is not consistent with $\Delta$. By contradiction, $\Delta$ is consistent.
\textbf{Qed. }

\subsection{Theorem \ref{theorem:pos_query}}
For a given specification configuration
$((\Sigma, \Phi), P, F, \Gamma_{P}, \Gamma_{F})$ and safe verification
interface $\Delta$, $\Delta(R_f)$ is weakest safe if and only if there are no weakening sample for $f$.

\textbf{Proof. }
We first prove if $\Delta(R_f)$ is weakest safe then there is no weakening sample. If there exists a weakening sample $s$, then $s$ includes a weakening feature vector $\I{fv}$. According to Def \ref{def:weakeningfv},
\begin{align*}
    \I{fv} \in \Delta(R_f)^- \text{ and } \Sigma[\Delta[R_f \mapsto \unitaryc{\I{fv}} \lor \Delta(R_f)]] \implies \Phi
\end{align*}
Then $\unitaryc{\I{fv}} \lor \Delta(R_f)$ is a weaker specification than $\Delta(R_f)$ and make $\Sigma[\Delta[R_f \mapsto \unitaryc{\I{fv}} \lor \Delta(R_f)]] \implies \Phi$. By contradiction, there is no weakening sample.

Then we prove if there are no weakening sample for $f$, $\Delta(R_f)$ is weakest safe. Assume there is no weakening sample but exists a weaker specification $\phi_f^w \not = \Delta(R_f)$ and makes $\Sigma[\Delta[R_f \mapsto \phi_f^w]] \implies \Phi$, According to Lemma \ref{lemma:specclassificatioin},
\begin{align*}
    \bigvee_{\I{fv} \in \phi_f^{w+}} \unitaryc{\I{fv}} &\not= \bigvee_{\I{fv} \in \Delta(R_f)^+} \unitaryc{\I{fv}} \\
    \bigvee_{\I{fv} \in \Delta(R_f)^+} \unitaryc{\I{fv}} &\implies \bigvee_{\I{fv} \in \phi_f^{w+}} \unitaryc{\I{fv}}  
\end{align*}
which means $\Delta(R_f)^+ \subset \phi_f^{w+}$, then there exists $\I{fv}$,
\begin{align*}
    &f \in \phi_f^{w+} \text{ and } \I{fv} \not\in \Delta(R_f)^+ \text{ and } \{\I{fv}\} \cup \Delta(R_f)^+ \subseteq \phi_f^{w+}
\end{align*}
As
\begin{align*}
&\Sigma[\Delta[R_f \mapsto \unitaryc{\I{fv}} \lor \Delta(R_f)]] \implies \Sigma[\Delta[R_f \mapsto \phi_f^w]]
\end{align*}
and
\begin{align*}
 \Sigma[\Delta[R_f \mapsto \phi_f^w]] \implies \Phi
\end{align*}
then
\begin{align*}
&\Sigma[\Delta[R_f \mapsto \unitaryc{\I{fv}} \lor \Delta(R_f)]] \implies \Phi
\end{align*}
which means $\I{fv}$ is a weakening feature vector. By contradiction, no weaker specification exists.
\textbf{Qed. }

\subsection{Theorem \ref{theorem:consistent-correct}}
Algorithm~\ref{algo:consistent} always halts; the verification interface returned by Algorithm~\ref{algo:consistent} is safe and consistent.


\textbf{Proof. } First we prove the decidability. Consider the size of positive feature vector set ($\pi_f$) and negative feature vector set ($\omega_f$) for each library function $f$. For the inner loop(line $4-10$), the algorithm always add new feature vector to $\pi_f$ on line $9$. According to the correctness of the property based sampler, for given $s^+$ generated by $\I{Sample}$ and each library function $f$, there always exists at least one feature vector extracted from $s^+$ not included by $\pi_f$. Thus the size of $\pi_f$ always increases, the total number of feature vectors of each library function is finite($2^{|\mathcal{S}|}$), the inner loop always halt. On the other hand, the algorithm always add new feature vectors to $\omega_f$(line $20$) in the outer loop. If the outer loop continues, the condition on line $19$ should be $\C{false}$, then there exists at least one library function whose negative feature vector set $\omega_f$ strictly increases. Again, since the total number of feature vectors of each library function is finite, we can conclude Algorithm~\ref{algo:consistent} always halts


The algorithm only terminates on line $16$ which returns verification interface $\Delta$. Notice that SMT solver proves there are no negative samples which is not consistent with $\Delta$ on line $12$ then according to Theorem \ref{theorem:safesolution}, $\Sigma[\Delta] \implies \Phi$. SMT solver also proves all $\Sigma[\Delta]$ is satisfiable on line $14$, then according to Definition~\ref{def:safe}, $\Delta$ is safe. On the other hand $\I{Sampler}$ on line $5$ shows all samples entailed by $\Gamma_F$ and $\Gamma_P$ are consistent with $\Delta$; according to Theorem \ref{theorem:safesolution}, $\Delta$ is consistent. Thus the algorithm returns a safe and consistent verification interface.
\textbf{Qed. }


\subsection{Theorem \ref{theorem:weakening-correct}}
For given $(\Sigma, \Phi), P, \Delta, f, \vec{u}$, if $\Delta$ is safe, algorithm~\ref{algo:single} will halt with the weakest safe specification for $f$.

\textbf{Proof. }
Now we first prove when algorithm halts with $W_f \lor \Delta(R_f)$ (or $\Delta(R_f)$, then this case $W_f \lor \Delta(R_f)$ is equal to $\Delta(R_f)$, we just unify all cases as $W_f \lor \Delta(R_f)$ in the proof) on line $5$, then the result is always weakest safe. The \autoref{algo:single} terminates on line $5$ which means $W_f \lor \Delta(R_f)$ is initially weakest safe or the algorithm has executed the $\I{SafetyLoop}$, then the SMT solver have proved
$\Sigma[\Delta[R_f \mapsto W_f \lor \Delta(R_f)]] \implies \Phi$(line $17$) is valid; as $\Delta$ is safe, notice that
\begin{align*}
    \Sigma[\Delta] \equiv \Sigma[\Delta[R_f \mapsto \Delta(R_f)]] \implies \Sigma[\Delta[R_f \mapsto W_f \lor \Delta(R_f)]]
\end{align*}
then
\begin{align*}
    \Sigma[\Delta[R_f \mapsto \Delta(R_f)]] \not\models False
\end{align*}
thus $\Delta[R_f \mapsto W_f \lor \Delta(R_f)]$ is also a safe verification interface. On the other hand, if $W_f \lor \Delta(R_f)$ is not weakest one, according to Theorem~\ref{theorem:pos_query}, there exists a weakening sample $s$. Then according to Def \ref{def:weakeningfv}, $s$ is positive with respect to $f$ that is $s \models \Sigma[\Delta[R_f \mapsto \top]]$; $s$ contains weakening feature vectors which not included by $W_f \lor \Delta(R_f)$ and any non-weakening feature vectors in $\omega$, thus
\begin{align*}
    s \models \Sigma[\Delta[R_f \mapsto \top]] \land \Phi \land \neg\Sigma[\Delta[R_f \mapsto \unitaryc{\omega} \lor W_f \lor \Delta(R_f)]]
\end{align*}
However, SMT solver have proved $s$ does not exists on line $3$. By contradiction, the result $W_f \lor \Delta(R_f)$ is weakest safe.

Then we prove Algorithm~\ref{algo:single} always halts. Consider the size of positive feature vector set($\pi$) and negative feature vector set($\omega$). $\pi$ and $\omega$ are always increased by $\I{Update}$, thus we can state $|\Pi| + |\Omega|$ \emph{never decreases} during the iterations of the loop. Moreover, consider the sample $s$ on line $3$,
\begin{align*}
    s \models \neg\Sigma[\Delta[R_f \mapsto \unitaryc{\omega} \lor W_f \lor \Delta(R_f)]]
\end{align*}
then
  \begin{align*}
    \exists R_f(\vec{\alpha}, \nu) \in \Sigma,\;
    \exists \mathit{fv} \in \chi_{\mathit{FSet}(P, f(\vec{\alpha}) = \nu, \vec{u})}(s),
    \mathit{fv}\not\in (\unitaryc{\omega} \lor W_f \lor \Delta(R_f))(\vec{a}, \nu)^+
  \end{align*}
there exists a feature vector extracted from $s$,
\begin{align*}
\exists \I{fv} \in \chi_{\mathcal{S}}(s), \I{fv} \not \in \omega \text{ and } \unitaryc{\I{fv}} \centernot\implies W_f \lor \Delta(R_f)
\end{align*}
As $W_f$ is the learning result from $\pi$ and $\omega$, thus according to Lemma \ref{lemma:specclassificatioin}, 
\begin{align*}
    \unitaryc{\pi} \implies W_f
\end{align*}
thus $\unitaryc{\I{fv}} \centernot\implies \unitaryc{\pi}$ which means $\I{fv}$ is not included in $\pi$. Then there is at least one feature vector not included in $\pi$ or $\omega$. Consider the sample $s$ on line $17$,
\begin{align*}
    s \models \Sigma[\Delta[R_f \mapsto W_f \lor \Delta(R_f)]]
\end{align*}
which is a negative sample inconsistent with $\Delta[R_f \mapsto W_f \lor \Delta(R_f)]$. By Def \ref{def:sampleconsistency} there exists a feature vector extracted from $s$ by not included in $\omega$. Moreover, if all feature vector extracted from $s$ are already included in $\pi$, thus $|\pi| + |\omega|$ always increases, we can conclude \autoref{algo:single} always halts.
\textbf{Qed. }

\subsection{Lemma Locally maximality}

\begin{lemma}[Locally maximality]
For given verification query $(\Sigma, \Phi)$, a consistent verification interface $\Delta$ is maximal with respect to $\vec{u}$ iff for each library function $f$, $\Delta(f)$ is weakest safe.
\label{lemma:local}
\end{lemma}
\textbf{Proof. }
If $\Delta$ is maximal but for one library function $f$, there exists a specification $\phi_f$ which is weaker than $\Delta(R_f)$ and $\Sigma[\Delta[R_f \mapsto \phi_f]]$ is still safe. By Def \ref{def:order}, $\Delta[R_f \mapsto \phi_f] \succ \Delta$ which is conflict with $\Delta$ is maximal. By contradiction, each specification in $\Delta$ is weakest safe. On the other hand, if each specification in $\Delta$ is weakest safe but $\Delta$ is not maximal, then there exists another verification interface $\Delta' \succ \Delta$. By Def \ref{def:order}, all specification in $\Delta'$ is weaker than corresponding specification in $\Delta$, and exists a library function $f$, $\Delta'(R_f) \not = \Delta(R_f)$. Then when we fixed the specification of functions besides $f$ to specification in $\Delta$, $\Delta'(R_f)$ is weaker than $\Delta(R_f)$ but still make the verification queries valid, which is conflict with $\Delta(R_f)$ is weakest safe. By contradiction, $\Delta$ is maximal verification interface.
\textbf{Qed. }

\subsection{Theorem \ref{theorem:multi-correct}}
Algorithm~\ref{algo:multi} either returns either a maximal verification interface or a concrete valued
counter-example. 

\textbf{Proof. } The algorithm terminates on line $6$ and $14$, returns a concrete valued counter-example; or a verification interface $\Delta$ literately weakened by $\I{Weaken}$ from the result from $\I{
SpecInfer}$, called $\Delta_{SpecInfer}$. According to Theorem \ref{theorem:consistent-correct}, $\Delta_{SpecInfer}$ is safe and consistent. Notice that $\Delta \succ \Delta_{SpecInfer}$, according to Lemma \ref{lemma:specclassificatioin}, all positive samples consistent with $\Delta_{SpecInfer}$ are also consistent with $\Delta$. As $\Delta_{SpecInfer}$ is consistent, by Theorem \ref{theorem:safesolution}, all positive samples entailed by $\Gamma_F$ and $\Gamma_P$ are consistent with $\Delta_{SpecInfer}$, thus also consistent with $\Delta$.  According to Lemma \ref{lemma:local} and Theorem \ref{theorem:weakening-correct}, each $\Delta(R_f)$ is weakest safe, then $\Delta$ is maximal verification interface.
\textbf{Qed. }

\section{Algorithm for a Set of Verification Queries}

\begin{algorithm}[t!]
  \Params{ Specification configuration $(\{(\Sigma, \Phi)\}, F, P, \Gamma_{F}, \Gamma_{P})$}
  \Output{ Maximal verification interface $\Delta$ or counterexample \textsc{Cex}}
  $\vec{u} \leftarrow \emptyset$\;
  \While{true}{
    \Match{$\I{SpecInfer}(\{(\Sigma, \Phi)\}, F, P, \Gamma_{F},
      \Gamma_{P}, \vec{u})$}{
      \Case{Fail $s^-$}{
        \Match{$\I{ExtractCex}(s^-)$}{
          \lCase{\textsc{Cex}}{
            \Return{\textsc{Cex}}
          }
          \lCase{None}{
            $\vec{u} \leftarrow \vec{u} \cup \I{FreshVariable}()$
          }
        }
      }
      \lCase{Fail None}{
        $\vec{u} \leftarrow \vec{u} \cup \I{FreshVariable}()$
      }
      \Case{$\Delta$}{
        \Repeat{$\Delta = \Delta_0$}{
          $\Delta_0 \leftarrow \Delta$\;
          \lFor{$f \in F$}{
            $\Delta(R_f) \leftarrow \I{Weaken}(\{(\Sigma, \Phi)\}, P, \Delta, f, \vec{u})$
          }
        }
        \Return{$\Delta$}\;
      }
    }
  }
  \caption{Multi-Abductive Inference Algorithm.}
  \label{algo:multi_full}
\end{algorithm}


The multi-abductive inference algorithm for multiple verification queries are shown in \autoref{algo:multi_full}, which is the same as \autoref{algo:multi} but takes a set of verification queries $\{(\Sigma, \Phi)\}$ as input.

\begin{algorithm}[hbt!]
\Params{Specification configuration $\{(\Sigma, \Phi)\}, F, P, \Gamma_{F},
  \Gamma_{P}$, variables in hypothesis space $\vec{u}$}
\Output{Consistent safe verification interface or reports failure}
\lFor{$f \in F$}{
               $\pi_f,\omega_f \leftarrow \emptyset, \emptyset$
          }
$\Delta \leftarrow \{R_f \mapsto \forall \vec{u}, \top\}$\;
\While{$\I{true}$}{
  \While{$true$}{
    \Match{$\I{Sample}(\Delta(R_f), \Gamma_{F}, \Gamma_{P})$}{
      \lCase{None}{\Break}
      \Case{Some $s^+$}{
        \For{$f \in F$}{
          $\pi_f \leftarrow
          \I{FvecFromSample}(\I{FSet}(P,f(\vec{\alpha}_f) = \nu_f , \vec{u}), s^+)~~ \cup~~ \pi_f$\;
          $\omega_f \leftarrow \omega_f \setminus \pi_f$}
        \lFor{$f \in F$}{
            $\Delta(R_f) \leftarrow \I{Learner}(\pi_f, \omega_f)$
        }
      }
    }
  }
  \If{\ZForall $(\Sigma_i, \Phi_i)$, $\I{Verify}((\Sigma_i[\Delta] \implies \Phi_i)) = \I{OK}$}
  {
      \leIf{\ZForall $(\Sigma_i, \Phi_i)$, $\I{Verify}((\Sigma_i[\Delta] \implies \Phi_i)) = \I{Sat}\ \_$}{
        \Return{$\Delta$}
      }{
        \Return{Fail None}
      }
  }
  \For{$(\Sigma_i, \Phi_i)$}{
      \Match{$\I{Verify}((\Sigma_i[\Delta] \implies \Phi_i))$}{
            \lCase{$\I{OK}$}{
             \Continue
            }
            \Case{$\I{Sat}\ s^-$}{
              \lFor{$f \in F$}{
                $\omega'_f \leftarrow
                \I{FVecFromSample}(\I{FSet}(P, f(\vec{\alpha}_f) = \nu_f , \vec{u}),
                s^-) ~~\setminus~~ \pi_f$
              }
              \lIf{$\bigwedge\limits_{f\in F} \omega'_f = \emptyset$}{
                \Return{Fail $s^-$}
              } \lElse {
              \ZFor $f \in F$ \ZDo
                $\omega_f \leftarrow \omega_f \cup \omega'_f$
              }
        }
      }
  }
                \lFor{$f \in F$}{
                $\Delta(R_f) \leftarrow \I{Learner}(\pi_f, \omega_f)$
          }
}

\caption{Safe and Consistent Specification Inference($\I{SpecInfer}$)}
\label{algo:consistent_full}
\end{algorithm}

The $\I{SpecInfer}$ for set of verification queries are shown in \autoref{algo:consistent_full} which has loops on lines $14 - 20$ which verifies each verification queries and gather negative feature vectors. When algorithm cannot gather new feature vectors in any verification queries on line $19$, then returns the negative sample. If all verification queries is valid and all $\Sigma_i[\Delta]$ is satisfiable, then returns the verification interface.

\begin{algorithm}[ht!]
\Params{$\{(\Sigma, \Phi)\}, P, \Delta, f, \vec{u}$}
\Output{Weakest safe specification of $f$}
$W_f \leftarrow \forall\vec{u}, \bot;\ \pi, \omega \leftarrow emptyset, \emptyset $\;
\While{$\I{true}$}{
    \Match{$\I{Verify}(\bigwedge (\Sigma_i[\Delta[R_f \mapsto \forall
      \vec{u}, \top]] \land \Phi_i \implies \Sigma_i[\Delta[R_f \mapsto
      \unitaryc{\omega} \lor W_f \lor \Delta(R_f)]]))$}
    {
      \uCase{$\I{OK}$}{
        \leIf{$W_f = \forall\vec{u}, \bot$}{\Return{$\Delta(R_f)$}}{\Return{$W_f \lor \Delta(R_f)$}}
      }
      \Case{$\I{Sat}\ s$}{
        $W_f, \pi, \omega \leftarrow $ $\I{Update}$ $(s, W_f, \pi, \omega)$\;
        $W_f, \pi, \omega \leftarrow $ $\I{SafetyLoop}$ $(W_f, \pi, \omega)$\;
      }
    }
  }

\Procedure{$\I{Update}(s, W_f, \pi, \omega)$}{
    \For{$\I{fv} \in \I{FVecFromModel}(Fset(P, f(\alpha_f) = \nu_f, \vec{u}),\; s)$}{
        $W_f \leftarrow \I{Learner}(\pi, \omega)$\;
        \Match{$\I{Verify}(\bigwedge
        (\Sigma_i[\Delta[R_f \mapsto W_f \lor \Delta(R_f) \lor \unitaryc{\I{fv}}]] \implies \Phi_i))$}{
            \lCase{$\I{OK}$}{
                $\pi \leftarrow \pi \cup \{\I{fv}\}$
            }
            \lCase{$\I{Sat}\ \_$}{
                $\omega \leftarrow \omega \cup \{\I{fv}\}$
            }
        }
    }
    \Return{$W_f, \pi, \omega$}
}

\Procedure{$\I{SafetyLoop}(W_f, \pi, \omega)$}{
    \Match{$\I{Verify}(\bigwedge(\Sigma_i[\Delta[R_f \mapsto W_f \lor \Delta(R_f)]] \implies \Phi_i))$}{
        \lCase{$\I{OK}$}{
            \Return{$W_f, \pi, \omega$}
        }
        \Case{$\I{Sat}\ s$}{
            $W_f, \pi, \omega \leftarrow \I{Update}(s, W_f, \pi, \omega)$\;
            \Return{$\I{SafetyLoop}(W_f, \pi, \omega)$}
        }
    }
}
\caption{Weakening Algorithm}
\label{algo:single_full}
\end{algorithm}

The weakening algorithm for set of verification queries are shown in \autoref{algo:single_full} where all SMT queries on line $3$, $12$ and $17$ are conjunctions of corresponding queries in \autoref{algo:single} for each verification query.
\else
\fi

\end{document}
\endinput